\documentclass[english,11pt]{article}

\usepackage{authblk}

\usepackage{relsize}
\usepackage{subcaption}
\usepackage[draft]{fixme}
\usepackage[T1]{fontenc}
\usepackage[latin9]{inputenc}
\usepackage[square]{natbib}
\usepackage[margin=1.25in]{geometry}
\usepackage{mathtools}
\usepackage{amsmath}
\usepackage{enumerate}
\usepackage{enumitem}
\DeclareFontFamily{U}{skulls}{}
\DeclareFontShape{U}{skulls}{m}{n}{ <-> skull }{}

\usepackage{amsthm}
\theoremstyle{plain}
\usepackage{amssymb}
\usepackage{calc}
\newlength{\myl}
\usepackage{setspace}
\usepackage{babel}
\usepackage{txfonts}

\usepackage{tcolorbox}
\usepackage{proof}

\usepackage{etoolbox,refcount}

\usepackage{multicol}

\newcounter{countitems}
\newcounter{nextitemizecount}
\newcommand{\setupcountitems}{%
  \stepcounter{nextitemizecount}%
  \setcounter{countitems}{0}%
  \preto\item{\stepcounter{countitems}}%
}
\makeatletter
\newcommand{\computecountitems}{%
  \edef\@currentlabel{\number\c@countitems}%
  \label{countitems@\number\numexpr\value{nextitemizecount}-1\relax}%
}
\newcommand{\nextitemizecount}{%
  \getrefnumber{countitems@\number\c@nextitemizecount}%
}
\newcommand{\previtemizecount}{%
  \getrefnumber{countitems@\number\numexpr\value{nextitemizecount}-1\relax}%
}
\makeatother    
\newenvironment{AutoMultiColItemize}{%
\ifnumcomp{\nextitemizecount}{>}{3}{\begin{multicols}{2}}{}%
\setupcountitems\begin{itemize}}%
{\end{itemize}%
\unskip\computecountitems\ifnumcomp{\previtemizecount}{>}{3}{\end{multicols}}{}}

\usepackage{lmodern}
\usepackage{blindtext}
\usepackage{framed}
\usepackage{graphicx}
\usepackage{xcolor}
\usepackage{url}
\usepackage{pgf}
\usepackage{tikz}
\usepackage[misc]{ifsym}
\usetikzlibrary{arrows,automata,calc}

\usepackage[colorlinks=true]{hyperref}%
\definecolor{dark-red}{rgb}{0.4,0.15,0.15}
\definecolor{dark-blue}{rgb}{0.15,0.15,0.5}
\definecolor{medium-blue}{rgb}{0,0,0.6}
\definecolor{blob}{RGB}{0,76,153}
\defcitealias{Schn:1971}{Ibid.}
\hypersetup{linktocpage,
    colorlinks, linkcolor={blob},
    citecolor={blob}, urlcolor={blob}
}

\newtheorem{thm}{Theorem}[section]

\newtheorem{lem}[thm]{Lemma}
\newtheorem{fact}[thm]{Fact}
\newtheorem{prop}[thm]{Proposition}
\newtheorem{defn}[thm]{Definition}

\newtheorem*{cla*}{Claim}

\newtheorem*{ope*}{Open problem}
\newtheorem{observation}[thm]{Observation}
\newtheorem{rem}[thm]{Remark}
\newtheorem*{rem*}{Remark}

\newcommand\blfootnote[1]{%
  \begingroup
  \renewcommand\thefootnote{}\footnote{#1}%
  \addtocounter{footnote}{-1}%
  \endgroup
}

\newcommand{\bang}[1]{\langle ! {#1}\rangle}

\newcommand{\del}[1]{\langle - {#1}\rangle}

\DeclareRobustCommand{\VAN}[3]{#3}

\usepackage{fancyhdr}

\fancypagestyle{firstpage}{%
  \lhead{}
  \rhead{Forthcoming in \emph{The Review of Symbolic Logic} \\(Cambridge University Press)\\ doi:10.1017/S1755020320000258 \\  \textcopyright  Association for Symbolic Logic 2020}
}

\fancypagestyle{norm}{%
  \lhead{}
  \rhead{Forthcoming in \emph{The Review of Symbolic Logic} \\(Cambridge University Press)\\ doi:10.1017/S1755020320000258 \\  \textcopyright  Association for Symbolic Logic 2020}
}

\title{The Modal Logic of Stepwise Removal}

\author[1,2,3]{Johan van Benthem}
\author[4]{Krzysztof Mierzewski}
\author[4]{Francesca Zaffora Blando}

\affil[1]{Stanford University and Logical Dynamics Lab, CSLI}
\affil[2]{ILLC, University of Amsterdam}
\affil[3]{Tsinghua University}
\affil[4]{Carnegie Mellon University}

\date{}

\begin{document}
\maketitle
\thispagestyle{firstpage}
\vspace{-4em}
\begin{abstract}
We investigate the modal logic of stepwise removal of objects, both for its intrinsic interest as a logic of quantification without replacement, and as a pilot study to better understand the complexity jumps between dynamic epistemic logics of model transformations and logics of freely chosen graph changes that get registered in a growing memory. After introducing this logic (\textsf{MLSR}) and its corresponding removal modality, we analyze its expressive power and prove a bisimulation characterization theorem. We then provide a complete Hilbert-style axiomatization for the logic of stepwise removal in a hybrid language enriched with nominals and public announcement operators. Next, we show that model-checking for \textsf{MLSR} is PSPACE-complete, while its satisfiability problem is undecidable. Lastly, we consider an issue of fine-structure: the expressive power gained by adding the stepwise removal modality to fragments of first-order logic. \\


\end{abstract}
\blfootnote{\Letter\,  Johan van Benthem: \texttt{j.vanbenthem@uva.nl}, Krzysztof Mierzewski: \texttt{kmierzew@andrew.cmu.edu}, Francesca Zaffora Blando: \texttt{fzaffora@andrew.cmu.edu}.}
\section{Model change and quantification}\label{Model change and quantification}

\noindent Logical systems describing model change come up when reasoning about forms of semantic interpretation that affect a current model, varieties of information update, or more general actions changing a local environment. A typical feature of such systems is the use of dynamic modalities that, when evaluated in a current model $\mathcal{M}$, look at what is true in other models $\mathcal{N}$, related to $\mathcal{M}$ via some relevant cross-model relation. These dynamic logics come in a wide range of expressive power and computational complexity \citep{AvBG:2018}. Our aim in this small pilot study is to explore a significant border line, where the complexity of the satisfiability problem jumps from decidable to undecidable. In the process, we highlight some further issues, as well as some new proof techniques, as will be explained below.  \\
\newgeometry{top=1.25in,bottom=1.25in}
\noindent\textbf{Dynamic epistemic logics of information update.} Here is one recent genre of dynamic logics that can describe model change. When modeling the effects of new information, a natural format changes a current epistemic model to a new one, suitably modified. For instance, an event $!\varphi$ of reliable public information that $\varphi$ is the case changes a current pointed model $(\mathcal{M}, s)$ to the definable sub-model $(\mathcal{M}| \varphi, s)$, whose domain is the set of all points in $\mathcal{M}$ that satisfy $\varphi$. Likewise, an event where all agents publicly lose all uncertainty about $\varphi$ takes $(\mathcal{M}, s)$ to a model $(\mathcal{M}\textbackslash \varphi, s)$, where the domain stays the same, but the epistemic accessibility relation $\sim$ of $\mathcal{M}$ gets replaced by the refinement $s \sim_\varphi t$: i.e., $s \sim t$ and, also, $\mathcal{M}, t\models \varphi$ if and only if $\mathcal{M}, s\models \varphi$. These and many other model transformations $F$ have matching modalities $[F]\psi$ in dynamic epistemic logics, whose key axioms for $[F]\psi$ give a recursive analysis of when the postconditions $\psi$ hold in terms of what was true before the $F$-update (see the survey by \cite{JB:2011}). Dynamic epistemic logics are usually decidable if their underlying static logics are: the recursion axioms reduce out the dynamic modalities, at least on full standard universes of epistemic models.\\

\noindent\textbf{Sabotage-style graph logics.} Here is a second natural genre of modal logics for describing model change. In the \emph{sabotage game} of \cite{vB:2005}, arbitrary links in a graph are cut, one by one, by a Demon opposing a Traveler, who, in turn, moves across the graph along still available links. The winning positions of the Demon and the Traveler can be analyzed using standard modalities, together with additional modalities describing what holds in a pointed model after one link has been removed from the current accessibility relation. However, validity in modal logics for various graph games of this sort can be undecidable, and the resulting model theory is quite complex (see \citep{AvBG:2018} and \citep{vBL:2018}).

\medskip

This difference in complexity calls for an explanation. The present paper locates its source in the contrast between, on the one hand, the simultaneous removal of points or links in dynamic epistemic logics and, on the other, the stepwise modifications captured by logics for sabotage and related graph games. In doing so, we explore the border between two system designs: dynamic epistemic logics of graph change that reduce effectively to a decidable static base language\textemdash and, hence, to what is true in the initial model, which already `pre-encodes' the effects of changes\textemdash and, on the other hand, undecidable sabotage-type logics of graph change operations, whose effects are not pre-encoded in the original model, but rather depend on a growing `memory' of previous changes.

To make this concrete, here is a simplest dynamic epistemic logic turned `stepwise'. For simplicity, we focus on point deletion, rather than link deletion.\\

\noindent\textbf{A stepwise update modality.} Consider the standard language of basic modal logic, augmented with a dynamic modality $\del{\varphi}\psi$ that has the following semantics. 
\begin{defn}\label{semantics}
Given a relational model $\mathcal{M}=(W, R, V)$, with $R\subseteq W\times W$ and $V$ a valuation, the satisfaction clause for $\del{\varphi}\psi$ reads 
$$
\mathcal{M}, s\models \langle-\varphi\rangle\psi  \text{ iff there is a point  $t \neq s$ in $\mathcal{M}$ with } \mathcal{M}, t \models \varphi \text{ and } \mathcal{M}-\{t\}, s\models\psi,
$$
where $\mathcal{M}-\{t\}$ is the submodel of $\mathcal{M}$ having just the point $t$ removed from its domain. More generally, given $D\subseteq W$, $\mathcal{M}-D$ denotes the submodel of $\mathcal{M}$ with domain $W\setminus D$.
\end{defn}
This system of what may be called stepwise point removal (\textsf{MLSR}) will be studied here as an intermediate case between the simplest dynamic epistemic logic of public announcements, where all points satisfying $\varphi$ are removed simultaneously during an update, and a simple sabotage modal logic for stepwise graph change.\\

\noindent\textbf{Quantification without replacement.} The language introduced here has various further interpretations. For instance, it can be seen as a medium for describing `interventions' that minimally change some given model to make some specified new properties true \citep{REN:2001}.  But the system has an even more general logical motivation, which is not tied to information updates or any other specific application. 

Consider the evaluation of restricted existential quantifiers $\exists x\, \varphi(x) \cdot \psi(x)$ in first-order logic (\textsf{FOL}). One searches for an object $d$ satisfying $\varphi$ and then checks whether  $d$ also satisfies $\psi$. In this second stage, the model has not changed: the witness $d$ is still in the domain and it influences the evaluation of $\psi$. Call this process ``quantification with replacement''. Now, it has been claimed \citep{HS:1997} that quantifiers in natural language can also behave differently: witness, for instance, the natural sense in which the distrust in ``John distrusted everyone'' does not apply to John himself. Even though this may be an idiosyncrasy of natural language, it clearly makes sense to explore quantification without replacement as a model for evaluation procedures that change domains \citep{DG:2013}:

\medskip

\indent $\exists x (\varphi|\psi)$ says that there is an object (or, in a natural polyadic version $\exists \overline{x} (\varphi|\psi)$, \\ 
\indent a tuple of objects) that satisfies $\varphi$ in the current model $\mathcal{M}$, while $\psi$ holds in the \\
\indent sub-model $\mathcal{M}-\{s\}$ where that object (or all those objects) has been removed.

\medskip

This quantifier form is clearly definable in \textsf{FOL} with identity, but, taken by itself, it suggests its own model theory and proof theory. Moreover, as we shall see, adding quantification without replacement to weaker fragments of the first-order language, such as monadic predicate logic or basic modal logic, produces much less simple effects.
\\

\noindent\textbf{The system \textsf{MLSR}.} The system \textsf{MLSR} of stepwise object removal studied in this paper provides a simple modal setting for bringing all of this out. Its syntax is that of the basic modal language with proposition letters, $\neg$, $\vee$, $\Diamond$, plus the additional modality $\langle-\varphi\rangle\psi$, whose semantics was given above (Definition \ref{semantics}). Occasionally, we will also use this language extended with a ``public announcement'', or relativization, modality $\langle!\varphi\rangle\psi$ describing what is true in restrictions to definable subdomains:
$$
\mathcal{M}, s\models \langle!\varphi\rangle\psi \;\; \text{ iff } \; \; \mathcal{M}, s\models\varphi \text{ and }\mathcal{M}|\varphi, s  \models \psi,
$$
with $\mathcal{M}|\varphi$ the submodel of $\mathcal{M}$ consisting of all and only the points in $\mathcal{M}$ where $\varphi$ is true. \\

\noindent\textbf{Outline of the paper.} In this paper, we study the essential features of this modal system. In \S\ref{Basics of expressive power}, we analyze the expressive power of \textsf{MLSR} by providing a first-order translation and a semantic characterization in terms of bisimulation invariance. This mainly requires straightforward adaptations of known techniques. \S\ref{Axiomatization} and \S\ref{Completness} present a complete axiomatization for \textsf{MLSR}, based on a new idea of mixing standard relativization with stepwise removal, which may very well be applicable to many other logics of graph change, for which Hilbert-style axiomatizations have long been an open problem. In \S\ref{Complexity and undecidability}, we first analyze the computational complexity of model checking for \textsf{MLSR}, which turns out to be PSPACE-complete. This analysis uses a reduction technique from \cite{LR:2003} which deserves to be better known in modal logic. Next, we prove that the satisfiability problem for \textsf{MLSR} is undecidable using a tiling argument familiar from the modal logic literature \citep{Marx:2006, AFH:2015}. In \S\ref{Stepwise removal over first-order fragments}, we then raise a more general definability issue: namely, what the addition of quantification without replacement does to various fragments of first-order logic. In particular, we  show that, when added to monadic first-order logic, the modality $\langle-\varphi\rangle\psi$ essentially allows us to count, boosting the expressive power of monadic first-order logic to that of monadic first-order logic with identity. 

In summary, we locate the threshold of complexity in the stepwise character of the modality for point removal, leading to the need for a computational device for maintaining a memory of deleted points, whose complexity equals that of arbitrary tiling problems and computations of Turing machines. In the process, we also raise new types of questions about modal logics of graph change, and we advertise and introduce some techniques that deserve to be better known among modal logicians.

\section{Basics of expressive power}\label{Basics of expressive power}

We start with the formal language to be used in most of this paper.\footnote{This language will be extended slightly with nominals in \S\ref{Axiomatization} and \S\ref{Completness}, which deal with proof systems.}

\begin{defn}\label{MLSR syntax}
The syntax of \emph{\textsf{MLSR}} is given by
$$
\varphi:= p\,|\,\neg\varphi\,|\,(\varphi\vee\varphi)\,|\,\Diamond\varphi\,|\,\langle-\varphi\rangle\varphi,
$$
with $p\in\emph{\textsf{PROP}}$. Dual modal operators $\Box, [-\varphi]$ are defined as usual.
\end{defn}

\paragraph{Some definable notions.} The language of \textsf{MLSR} can define various modal operators from hybrid logic \citep{AtC:2006} that go beyond the basic modal language. For instance, the difference modality $\mathsf{D}\varphi$ (`$\varphi$ is true at some different point') can be defined as $\langle-\varphi\rangle\top$, and this, in turn, allows to define the existential modality $\mathsf{E}\varphi$ as $\varphi\vee \mathsf{D}\varphi$. 
\textsf{MLSR} can also count all finite cardinalities, using suitably iterated formulas $$\underbrace{\langle-\top\rangle...\langle-\top\rangle}_\text{$k$ times}\top,$$ 
which express that a model has at least $k$ objects different from the current point of evaluation.  In addition, \textsf{MLSR} can define quite a few finite relational graphs up to isomorphism. For instance, let $\rho_{2}$ be the formula defining domain size 2, and let $\mathsf{U}$ be the universal modality (i.e., $\mathsf{U}\varphi = \varphi\wedge [-\neg\varphi]\bot)$. The following observation requires an easy exercise in understanding what our language can express.

\begin{fact}
The \emph{\textsf{MLSR}}-formula $\rho_{2} \wedge \mathsf{U}\langle-\top\rangle\Box\bot \wedge \Diamond\Diamond\top$ defines a two-point irreflexive loop.
\end{fact}

However, not every finite graph is definable, as we shall soon see.\\

\noindent\textbf{\textsf{SR}-bisimulation.} The semantic invariance matching this language is as follows.

\begin{defn}
A relation $Z$ between a set of pointed relational models is an \emph{\textsf{SR}}-bisimulation if it is a modal bisimulation in the ordinary sense, where the back and forth clauses stay inside the same models $\mathcal{M}, \mathcal{N}$, while, in addition, 
\begin{itemize}
\item[\emph{(a)}] if $(\mathcal{M}, s) Z (\mathcal{N}, t)$ and $u \in \mathcal{M}$ with $u \neq s$, then there is a $v \in \mathcal{N}$ such that $v \neq t$, $(\mathcal{M},u) Z (\mathcal{N}, v)$, and $(\mathcal{M}-\{u\},s) Z (\mathcal{N}-\{v\}, t)$, 
\item[\emph{(b)}] the analogous clause in the converse direction.
\end{itemize}
\end{defn}

Note that this definition imposes some minimal closure conditions on the set of models involved in the above clauses that are easy to spell out. The following property is proved by a standard induction on formulas.

\begin{fact}
\emph{\textsf{MLSR}}-formulas are invariant for \emph{\textsf{SR}}-bisimulations.
\end{fact}

Now we can give an example of two finite graphs that are not definable up to isomorphism and, in line with this, a first-order formula that is not in \textsf{MLSR}.

\begin{fact}\label{not LSR-def}
$\forall y (Rxy\vee Ryx)$ is not \emph{\textsf{MLSR}}-definable.
\end{fact}
\begin{proof}
Consider the model $\mathcal{M}$ consisting of two isolated reflexive points and the model $\mathcal{N}$ consisting of two points with the universal relation, plus all their submodels. By checking all clauses, one sees that the universal relation $Z$ between all pairs $(\mathcal{M}, x)$ and $(\mathcal{N}, y)$ plus all links between the 1-point pointed sub-models of $\mathcal{M}$ and $\mathcal{N}$ is an \textsf{SR}-bisimulation. But, clearly, connectedness holds in $\mathcal{N}$, but not in $\mathcal{M}$.  	              
\end{proof}

This new logical system still lies inside standard first-order logic.
\begin{fact}\label{translation}
There is an effective meaning-preserving translation from \emph{\textsf{MLSR}} into \emph{\textsf{FOL}}.
\end{fact}
\begin{proof}
We define the following compositional translation $\tau(\varphi, y, X)$ from \textsf{MLSR}-formulas $\varphi$ to first-order formulas, where $y$ is a free variable and $X$ a finite set of variables:
\begin{align*}
\tau(p, y, X)  &=  Py,\\
\tau(\neg\varphi, y, X)  &=  \neg\tau(\varphi, y, X),\\    \tau(\varphi\vee\psi, y, X)  &=  \tau(\varphi , y, X) \vee \tau(\psi, y, X),\\  
\tau(\Diamond\varphi, y, X)   &=  \exists z \bigg(Ryz \wedge \bigwedge_{x\in X}\neg(z = x) \wedge \tau(\varphi, z, X)\bigg),\\    
\tau(\langle-\varphi\rangle\psi, y, X) &=   \exists z \bigg(\neg(z=y) \wedge \bigwedge_{x\in X}\neg(z = x) \wedge \tau(\varphi, z, X) \wedge  \tau(\psi, y, X \cup \{z\})\bigg).
\end{align*}
Let $(\mathcal{M}, s)$ be any pointed model and $D = \{d_1,..., d_k\}$ a finite set of points in $\mathcal{M}$ of size $k$. The following equivalence is shown by a straightforward induction on \textsf{MLSR}-formulas $\varphi$ and sets of variables $X = \{x_1,..., x_k\}$ of size $k$:
$$
	\mathcal{M}- D, s \models \varphi   \text{ iff }   \mathcal{M}, a[y/s, X/D] \models  \tau(\varphi, y, X), 	
$$
where $a[y/s, X/D]$ is the variant of the variable assignment $a$ such that $a[y/s, X/D](y)=s$ and $a[y/s, X/D](x_i) = d_i$ for $1\leq i\leq k$. As a special case, there is an equivalence for \textsf{MLSR}-formulas in ordinary relational models $\mathcal{M}$ with $D = \varnothing$. 	
\end{proof}
\begin{rem}
The set $X$ in this translation serves as a finite memory storing the points that have already been deleted. This is an essential difference with first-order translations for standard modal languages, which usually lie inside fixed finite-variable fragments.
\end{rem}

A simple adaptation of a well-known model-theoretic argument for standard modal logic (cf. \citep{BRV:2001}) yields the following result.
\begin{thm}
The following assertions are equivalent for all first-order formulas $\varphi(x)$ in the signature of our models, with one free variable:
\begin{itemize}
    \item[\emph{(a)}] $\varphi(x)$ is invariant for \emph{\textsf{SR}}-bisimulation;
        \item[\emph{(b)}] $\varphi(x)$ is equivalent to the translation of some \emph{\textsf{MLSR}}-formula.
\end{itemize}
\end{thm}
\begin{proof}
We merely outline the points that need attention in the non-trivial direction from (a) to (b). Let $\mathcal{SR}$ denote the $\mathsf{MLSR}$-fragment of first-order logic (that is, all first-order formulas equivalent to translations of $\mathsf{MLSR}$ formulas via the translation $\tau$ from Fact \ref{translation}). 
As usual, one shows that $\varphi(x)$ is a semantic consequence of the set $\mathcal{C}_{x}(\varphi)$ of its $\mathcal{SR}$-consequences and then applies Compactness to get an $\mathcal{SR}$-equivalent. We thus need to show that $\mathcal{C}_{x}(\varphi) \models \varphi(x)$. Suppose $\mathcal{M}, s\models \mathcal{C}_{x}(\varphi)$. A standard compactness argument shows that there is a model $\mathcal{N}$ and $t\in\mathcal{N}$ such that $(\mathcal{M}, s)$ and $(\mathcal{N}, t)$ are $\mathcal{SR}$-equivalent, while $\mathcal{N}, t \models\varphi(x)$. These models are then extended to $\omega$-saturated elementary extensions $(\mathcal{M}^+, s)$ and $(\mathcal{N}^+, t)$. We use first-order saturation allowing finite sets of parameters consisting of designated objects in the models; in turn, the finitely satisfiable sets of first-order formulas to be saturated can have a finite set of free variables (not just one, as in the argument for basic modal logic). This is needed for the saturation argument to follow. 

Now we define a relation $Z$ between pointed models $(\mathcal{M}^+-D, u)$ and $(\mathcal{N}^+-E, v)$, with $E, D$ of the same finite size, which holds if $(\mathcal{M}^+-D, u)$ and $(\mathcal{N}^+-E, v)$ satisfy the same $\mathcal{SR}$-formulas. Using saturation, it can be shown that $Z$ is an \textsf{SR}-bisimulation, where the argument for the modality $\Diamond\varphi$ is standard, while the one for $\langle-\varphi\rangle\psi$ in terms of removing single objects goes as follows. Take $(\mathcal{M}^+-D, u)$ and $w \neq u$. Now, let 
\begin{align*}
  \Gamma(y) &:=  \big\{\gamma(y)\in\mathcal{SR}\,\big|\, \mathcal{M}^{+}-D, w\models \gamma\big\} \\
  \Delta(x) &:= \big\{\delta(x)\in\mathcal{SR}\,\big|\, \mathcal{M}^{+}-(D\cup\{w\}), u\models\delta\} 
\end{align*}
and consider the set of first-order formulas
$$
p(x,y)	:=\{\neg(y = x)\} \cup  \Gamma(y)\cup \Delta(x)
$$
This set is finitely satisfiable in $(\mathcal{M}^+-D, u, w)$ (interpreting $x$ as $u$ and $y$ as $w$). For each of its finite subsets $\{\neg(y = x)\} \cup  \Gamma^{\prime}(y)\cup \Delta^{\prime}(x)$, we have 
$$
\mathcal{M}^+-D, u, w\models \neg(y = x)\wedge \bigwedge \Gamma^{\prime}(y) \wedge \bigwedge \Delta^{\prime}(x),
$$
which means that 
$$
\mathcal{M}^+-D, u\models\exists y\Big( \neg(y = x)\wedge \bigwedge \Gamma^{\prime}(y) \wedge \bigwedge \Delta^{\prime}(x)\Big),
$$
and this formula is in $\mathcal{SR}$ (it is equivalent to the translation of a $\langle-\varphi\rangle\psi$ formula). This means that the formula also holds in $(\mathcal{N}^+-E, v)$. Thus, every finite subset of $p(x,y)$ is satisfiable in $(\mathcal{N}^+-E, v)$ (interpreting $x$ as $v$). In other words, expanding the language with a new constant symbol $\mathbf{c}$, the 1-type $p(\mathbf{c},y)$ is finitely satisfiable in $(\mathcal{N}^+-E, v)$ (fixing the interpretation of $\mathbf{c}$ as $v$).  Then, by saturation, the type  is realized in $(\mathcal{N}^+-E, v)$: we can thus find an object in $\mathcal{N}^+-E$ matching the given $w$, as required for an \textsf{SR}-bisimulation. 	
\end{proof}

\begin{rem}The first-order translation for \emph{\textsf{MLSR}} can also be phrased in terms of the hybrid language \emph{$H(E,\downarrow)$}, \emph{\citep{AtC:2006}}. The key translation clause here reads, for each formula of the form $\del{\varphi}\psi$ and sequence of nominals $\overline{\mathbf{n}}=(\mathsf{n}_{1},...,\mathsf{n}_{\ell})$:
$$
\sigma(\del{\varphi}\psi)^{\overline{\mathbf{n}}} = \downarrow_{\mathsf{m}}. \mathsf{E}\downarrow_{\mathsf{k}}. \bigg(\neg \mathsf{m} \wedge\bigwedge^{\ell}_{i=1}\neg\mathsf{n}_{i} \wedge \sigma(\varphi)^{\overline{\mathbf{n}}}\wedge @_{\mathsf{m}}\sigma(\psi)^{\overline{\mathbf{n}}, \mathsf{k}}\bigg)
$$
Further connections of \emph{\textsf{MLSR}} with hybrid logics will be discussed in \S\ref{Conclusion} below.
\end{rem}

\section{Axiomatization}\label{Axiomatization}

Thanks to the first-order translation, the valid formulas of \textsf{MLSR} are effectively axiomatizable. But more immediate information comes from explicit modal laws. For instance, the removal modality $\langle-\varphi\rangle\psi$ distributes over disjunction in both of its arguments:
\begin{fact}
The following formulas are both valid: 
\begin{align*}
\langle -\psi\rangle (\varphi_{1}\vee\varphi_{2}) &\leftrightarrow \big(\langle -\psi\rangle\varphi_{1} \vee \langle -\psi\rangle\varphi_{2}\big)
 \\
  \langle -(\varphi_{1}\vee\varphi_{2})\rangle\psi &\leftrightarrow \big(\langle -\varphi_{1}\rangle\psi \vee \langle -\varphi_{2}\rangle\psi\big)  
\end{align*}\label{VDISTR}
\end{fact}

\vspace{-7mm}

To obtain an explicit modal axiomatization, we extend the language of \textsf{MLSR} with a countable set \textsf{NOM} of nominals, each standing for either a unique point in the model, or not denoting at all (this small technical deviation from hybrid logic will be helpful later on.) We also add standard public announcement modalities $\langle !\varphi\rangle\psi$ from dynamic epistemic logic, whose interpretation was given in \S\ref{Model change and quantification}. This will turn out to be useful, even though the axiom system to follow features no recursion axioms in the usual dynamic epistemic style for the removal modality. For simplicity, we retain the name \textsf{MLSR} for this logic.

\begin{rem}
 There seem to be no modal recursion axioms inverting the operator order for combinations $\del{\varphi}\bang{\alpha}\psi$ or $\bang{\alpha}\del{\varphi}\psi$. For example, $\bang{\alpha}\del{\varphi}\psi$ is not equivalent to $\alpha\wedge\del{\bang{\alpha}\varphi}\bang{\alpha}\psi$ (consider, for instance, the case where $\alpha=\Diamond p$, $\varphi=\Box\bot$ and $\psi = \top$). This feature of the modal language may be contrasted with how first-order logic augmented with an explicit syntactic operator of relativization would write this recursion:  
$$(\exists x (\varphi|\psi))^{\alpha(\cdot)} (x) \, \leftrightarrow \, \alpha(x) \wedge \exists y(\alpha(y)\wedge y\neq x \wedge \varphi^{\alpha(\cdot)}(y) \wedge (\psi)^{\alpha(\cdot)\wedge\, \cdot\,\neq (y)}(x))$$ 
\end{rem}
 We now extend the language of Definition \ref{MLSR syntax} with nominals, public announcement operators, as well as the existential modality. 
\begin{defn}
\emph{\textsf{MLSR}} \emph{with nominals} (for short still to be called \emph{\textsf{MLSR}}) has the syntax
$$
\varphi:= p\,|\,\mathsf{n}\,|\,\top\,|\,\neg\varphi\,|\,(\varphi\vee\varphi)\,|\,\Diamond\varphi\,|\,\langle!\varphi\rangle\varphi\,|\,\langle-\varphi\rangle\varphi\,|\,\mathsf{E}\varphi,
$$
with $p\in\emph{\textsf{PROP}}$,  $\mathsf{n}\in\emph{\textsf{NOM}}$. Dual modal operators $\Box, [!\varphi], [-\varphi]$ and $\mathsf{U}$ are defined as usual.
\end{defn}

Note that it is not necessary to add the $@_\mathsf{n}$ operator from hybrid logic as a primitive symbol, for it can be defined using the universal modality: in our setting with (possibly non-referring) nominals, $@_{\mathsf{n}}\varphi$ is simply a shorthand for $\textsf{U}(\mathsf{n}\rightarrow \varphi)$.\footnote{For a further study of combining dynamic epistemic proof systems with hybrid logic, see \citep{JUH:2011}.} The following proof system may look somewhat complex, but its components just follow the formal syntax just introduced.

\begin{figure}
\begin{tcolorbox}[width=\textwidth]
\begin{minipage}{\textwidth}
\begin{center}
\textbf{The System }$\mathsf{MLSR}$ \\\hrulefill \\
\end{center}
\begin{itemize}
\item The rule of Replacement of Equivalents:
$$
\infer[(\textbf{RE})]{\alpha(\varphi) \leftrightarrow \alpha[\psi/ \varphi]}{\varphi\leftrightarrow\psi}
$$
\item All tautologies of classical propositional logic, plus the Modus Ponens rule
\item  Modal \textsf{K} axioms and rules for all universal modalities $\Box$, $\mathsf{U}$, $[!\varphi]$ and $[-\varphi]$ 
\item S5-axioms for the universal modality $\mathsf{U}$, plus the axiom $\mathsf{U}\varphi \rightarrow \Box\varphi$
 
\item Axioms for \textsf{PAL}:
 \begin{AutoMultiColItemize}
    \item[] $\bang{\varphi}p\leftrightarrow \varphi\wedge p$ ($p\in\mathsf{PROP}$)
    \item[] $\bang{\varphi}\mathsf{n}\leftrightarrow \varphi\wedge \mathsf{n}$ ($\mathsf{n}\in\mathsf{NOM}$)
    \item[] $\bang{\varphi}\top \leftrightarrow \varphi$
    \item []$\bang{\varphi}\neg\psi \leftrightarrow (\varphi \wedge \neg  \bang{\varphi}\psi)$
    \item[] $\bang{\varphi} (\psi \vee \alpha) \leftrightarrow (\bang{\varphi}\psi \vee \bang{\varphi} \alpha) $
    \item[] $\bang{\varphi}\Diamond \psi\leftrightarrow (\varphi \wedge \Diamond \bang{\varphi}\psi) $
    \item[] $\bang{\varphi}\bang{\psi}\alpha \leftrightarrow \bang{(\varphi\wedge [!\varphi]\psi)}\alpha$
    \item []$\bang{\varphi}\mathsf{E}\psi\leftrightarrow (\varphi\wedge\mathsf{E}\bang{\varphi}\psi)$
 \end{AutoMultiColItemize}
 
 \vspace{-1mm}
 
\item The Truth Axiom: $\bang{\top}\varphi \leftrightarrow \varphi$
\item Hybrid axiom:
 \begin{AutoMultiColItemize}
    \item[](\textbf{H})   $\mathsf{E} (\mathsf{n}\wedge \varphi)\rightarrow \mathsf{U}(\mathsf{n}\rightarrow \varphi)$
 \end{AutoMultiColItemize}

Hybrid inference rules:
$$
\qquad
\infer [(\mathsf{m}\not\in\varphi)\,\,(\textbf{Name})]{\varphi}{\mathsf{m}\rightarrow \varphi}
$$
$$
\infer[(\mathsf{m}\not\in \varphi,\sigma \text{ and } \nabla\in\{\Diamond,\mathsf{E}\})\,\,(\textbf{Paste})]{\mathsf{E}(\mathsf{n}\land \nabla \varphi)\rightarrow\sigma}{\big(\mathsf{E}(\mathsf{n}\land \nabla \mathsf{m}) \wedge \mathsf{E}(\mathsf{m}\land \varphi)\big)\rightarrow \sigma }
$$

\item Axiom for the removal modality:

\smallskip
\quad \quad (\textbf{Mix}) $(\mathsf{E}(\mathsf{n} \wedge \alpha) \wedge \langle!\neg\mathsf{n} \rangle\varphi) \rightarrow \langle - \alpha\rangle\varphi$
\end{itemize}
\quad \quad \, Inference rule for the removal modality:
$$
\infer[(\mathsf{k}\not\in \varphi,\alpha,\psi,\sigma)\,\,(\textbf{Mix Rule})] {\mathsf{E}\big(\mathsf{n}\wedge \bang{\varphi}\langle-\alpha\rangle\psi\big)\rightarrow\sigma}{\mathsf{E}\big(\mathsf{n}\wedge \bang{\varphi} (\mathsf{E}(\mathsf{k}\wedge\alpha)\wedge \bang{\neg\mathsf{k}}\psi)\big)\rightarrow\sigma}
$$
\vspace{-0.2em}
\end{minipage}\\
\end{tcolorbox}
\caption{The Hilbert-style proof system for \textsf{MLSR}.}\label{AXIOMS}
\end{figure}

\vspace{-2mm}

\begin{defn}
The logic \emph{\textsf{MLSR}} (see \emph{Figure \ref{AXIOMS}}) consists of:
\begin{itemize}
\item the rule of \emph{Replacement of Provable Equivalents},\footnote{This rule is the basis for any ordinary logical system. In particular, in   $\mathsf{MLSR}$, it applies to formulas following modalities as well as formulas occurring inside announcement and deletion modalities.}
    \item the axioms and rules of classical propositional logic;
    \item the axioms and rules of the minimal normal modal logic for all the universal box modalities of the language (static or dynamic), plus the standard axioms and rules for the global universal modality \emph{\citep{BRV:2001}};
     \item the \emph{Name Rule} and the \emph{Paste Rule} from hybrid logic \emph{\citep{AtC:2006}}, with the latter slightly adapted to our setting;
     \item the axiom $\mathsf{E} (\mathsf{n}\wedge \varphi)\rightarrow \mathsf{U}(\mathsf{n}\rightarrow \varphi)$, which we denote by \emph{(H)};
     \item  the usual reduction axioms of public announcement logic \emph{\textsf{PAL}} for atoms (including nominals), the existential base modality,  the global existential modality,  and the announcement modality    \emph{\citep{JB:2011}},\footnote{A reduction axiom for disjunction is supplied by the minimal modal logic for announcement modalities.} as well as the \emph{Truth Axiom} $\langle!\top\rangle\varphi \leftrightarrow \varphi$; 
   
    \item the following two principles connecting the stepwise removal modality with the public announcement modality:
\begin{enumerate}[align=left]
\item[\emph{(Mix Axiom)}] $(\mathsf{E}(\mathsf{n} \wedge \alpha) \wedge \langle!\neg\mathsf{n} \rangle\varphi) \rightarrow \langle - \alpha\rangle\varphi$;
\smallskip
\item[\emph{(Mix Rule)}] If \,  $\vdash 
\mathsf{E}(\mathsf{n}\land \bang{\varphi}\mathsf{E}(\mathsf{k}\land\alpha) \land \bang{\varphi}\bang{\neg\mathsf{k}}\psi)\rightarrow\sigma$,

\smallskip

then  $\vdash
\mathsf{E}(\mathsf{n}\land\bang{\varphi}\langle-\alpha\rangle\psi)\rightarrow\sigma$, \, where $\mathsf{k}\notin\sigma,\varphi,\alpha,\psi$.  
\end{enumerate}
\end{itemize}
\end{defn}

\begin{fact}
The \emph{Mix Axiom} is valid, and the \emph{Mix Rule} is semantically sound.
\end{fact}

\begin{rem}\label{weak hybrid}
The system \emph{\textsf{MLSR}} does not include all the usual axioms for the basic hybrid language because nominals can fail to denote in our models after an update. In particular, after the deletion of a state named by $\mathsf{n}$, the formula  $\neg\mathsf{E}\mathsf{n}$ holds. Connected to this, the equivalence $\mathsf{E}(\mathsf{n} \land \neg\varphi) \leftrightarrow \neg\mathsf{E}(\mathsf{n} \land \varphi)$  underpinning the common hybrid notation $@_\mathsf{n}$ is no longer valid.  However, the proof principles of \emph{\textsf{MLSR}} guarantee all the properties of nominals that we need in what follows. In particular, the following useful facts are provable:
\begin{itemize}
\item $\mathsf{E}(\mathsf{n} \land \neg\varphi) \leftrightarrow (\mathsf{E}\mathsf{n} \land \neg\mathsf{E}(\mathsf{n} \land \varphi))$
 \item   $\mathsf{n} \rightarrow (\mathsf{E}(\mathsf{n}\land\varphi)\leftrightarrow \varphi)$
    \end{itemize}
\end{rem}

The language of \textsf{MLSR}  captures various global properties of our semantics, such as the fact that nominals hold at one state at most. Deriving this  shows the Mix Rule at work.

\begin{observation}
The formula $\mathsf{n}\rightarrow \neg\del{\mathsf{n}}\top$ is an $\mathsf{MLSR}$ theorem for any $\mathsf{n}\in\mathsf{NOM}$.
\end{observation}
\begin{proof}
Take $\varphi, \psi = \top$,  $\sigma = \bot$,    $\alpha = \mathsf{n}$. Then the antecedent formula in the Mix Rule reads

$$ \mathsf{E}(\mathsf{n} \land \langle!\top\rangle \mathsf{E}(\mathsf{k}\land \mathsf{n}) \land \langle!\top\rangle\langle!\neg \mathsf{k}\rangle\top) \rightarrow \bot$$

\vspace{2mm}

\noindent This is derivable in $\mathsf{MLSR}$. Using Replacement of Equivalents,\footnote{This basic rule of our proof system will be appealed to tacitly at many places in what follows.} and appealing to (i) a simple analysis of $\langle!\top\rangle \mathsf{E}(\mathsf{k}\land \mathsf{n})$ using  the  $\mathsf{PAL}$ reduction axioms for $\mathsf{E}$ and nominals, 
and (ii) the implication from  $\langle!\neg \mathsf{k}\rangle\top$ to $\neg\mathsf{k}$ which is one half of the $\mathsf{PAL}$ reduction axiom for $\top$, the antecedent of the above formula derives $\mathsf{E}(\mathsf{k}\wedge \neg\mathsf{k})$. It then suffices to note that the S5 axioms for quantifiers allow to derive $\mathsf{E}(\mathsf{k}\wedge \neg\mathsf{k})\rightarrow \bot$.

Therefore, the consequent formula is provable using the Mix Rule:

$$ \mathsf{E}(\mathsf{n} \land \langle!\top\rangle\langle -\mathsf{n}\rangle\top) \rightarrow \bot$$

\vspace{2mm}

\noindent Using the Truth Axiom and the S5 axioms for quantifiers, this is equivalent in $\mathsf{MLSR}$ to $\mathsf{U}(\mathsf{n}\rightarrow\neg\del{\mathsf{n}}\top)$, which implies the desired $\mathsf{n}\rightarrow \neg\del{\mathsf{n}}\top$.\end{proof}
To increase familiarity with the proof system, we explore  $\mathsf{MLSR}$ a bit further.
\begin{rem}
\begin{itemize}
\item[\emph{(a)}] Here is a more elaborate derivation showing the interplay of the two dynamic modalities. The premise of the above Mix Rule uses antecedents prefixed by an existential modality. However, we can also derive the following `bare' variant:

$$
\infer[(\mathsf{k}\not\in \varphi,\alpha,\psi,\sigma)\,\,\quad (\emph{\textbf{Stripped Mix Rule}})] { \bang{\varphi}\langle-\alpha\rangle\psi\rightarrow\sigma}{  \bang{\varphi}(\mathsf{E}(\mathsf{k}\wedge\alpha)\wedge \bang{\neg\mathsf{k}}\psi)\rightarrow\sigma}
$$
To see this, assume the premise. Take a fresh nominal $\mathsf{n}$, and using propositional logic, derive
$(\mathsf{n} \land \bang{\varphi}(\mathsf{E}(\mathsf{k}\wedge\alpha)\wedge \bang{\neg\mathsf{k}}\psi))\rightarrow\sigma$. Given the facts derived in Remark \ref{weak hybrid}, this is equivalent to $(\mathsf{n} \land \mathsf{E}(\mathsf{n} \land  \bang{\varphi}(\mathsf{E}(\mathsf{k}\wedge\alpha)\wedge \bang{\neg\mathsf{k}}\psi)))\rightarrow\sigma$. Again by propositional logic, this yields $\mathsf{E}(\mathsf{n} \land  \bang{\varphi}(\mathsf{E}(\mathsf{k}\wedge\alpha)\wedge \bang{\neg\mathsf{k}}\psi))) \rightarrow (\mathsf{n} \rightarrow \sigma)$. Here, since  $\mathsf{n}$ was fresh, the nominal $\mathsf{k}$ still satisfies the conditions of the Mix Rule. Therefore, we can conclude $\mathsf{E}(\mathsf{n} \land  \bang{\varphi}\langle-\alpha\rangle\psi)\rightarrow (\mathsf{n} \rightarrow \sigma)$. From this,  using propositional logic,  $(\mathsf{n} \land \mathsf{E}(\mathsf{n} \land  \bang{\varphi}\langle-\alpha\rangle\psi))\rightarrow \sigma$. Then using Remark \ref{weak hybrid} once more, we get $(\mathsf{n} \land   \bang{\varphi}\langle-\alpha\rangle\psi)\rightarrow \sigma$, and with propositional logic, $\mathsf{n} \rightarrow   (\bang{\varphi}\langle-\alpha\rangle\psi)\rightarrow \sigma)$. Finally, using the Name Rule of the hybrid logic component of $\mathsf{MLSR}$, the  conclusion $\bang{\varphi}\langle-\alpha\rangle\psi \rightarrow \sigma$ follows. 

Taking the special case of $\varphi = \top$, and using the Truth Axiom of $\mathsf{MLSR}$ (which was not used in the preceding derivations), the Stripped Mix Rule reduces to:
$$
\infer[(\mathsf{k}\not\in \alpha,\psi,\sigma)\,\,\quad (\emph{\textbf{Basic Mix Rule}})] { \langle-\alpha\rangle\psi\rightarrow\sigma}{  (\mathsf{E}(\mathsf{k}\wedge\alpha)\wedge \bang{\neg\mathsf{k}}\psi)\rightarrow\sigma}
$$

\vspace{-3mm}

\item[\emph{(b)}] \emph{\textsf{MLSR}} also admits the following simple variant of the Paste Rule:
$$
\infer[(\mathsf{k}\not\in \varphi,\sigma)\,\,\quad (\emph{\textbf{Basic Paste Rule}})] {\mathsf{E}\varphi\rightarrow \sigma }{\mathsf{E}(\mathsf{k}\wedge \varphi)\rightarrow \sigma }
$$
The Basic Paste Rule is derivable by the preceding method, starting with the premise:
\begin{align*}
    &\vdash \mathsf{E}(\mathsf{k}\wedge \varphi)\rightarrow \sigma \\
    & \vdash \mathsf{E}(\mathsf{k}\wedge \varphi)\rightarrow (\mathsf{n}\rightarrow \sigma)\,\,\, (\text{by propositional logic; where $\mathsf{n}$ is a fresh nominal})\\
     & \vdash \big(\mathsf{E}(\mathsf{n} \wedge \mathsf{E}\mathsf{k})\wedge \mathsf{E}(\mathsf{k}\wedge \varphi)\big) \rightarrow (\mathsf{n}\rightarrow \sigma)\,\,\,(\text{by propositional logic})\\
      & \vdash \mathsf{E}(\mathsf{n} \wedge \mathsf{E} \varphi) \rightarrow (\mathsf{n}\rightarrow \sigma)\,\,\,(\text{by the Paste Rule})\\
        & \vdash \big(\mathsf{n}\wedge \mathsf{E}(\mathsf{n} \wedge \mathsf{E} \varphi)\big) \rightarrow  \sigma \,\, (\text{by propositional logic})\\
           & \vdash (\mathsf{n}\wedge \mathsf{E}\varphi) \rightarrow  \sigma\,\,\,\,(\text{by Remark \ref{weak hybrid} and Replacement of Equivalents)}\\
           & \vdash \mathsf{n}\rightarrow  (\mathsf{E}\varphi \rightarrow  \sigma)\,\, (\text{by propositional logic})\\
            & \vdash \mathsf{E}\varphi \rightarrow  \sigma\,\,\,(\text{by the Name Rule, since $\mathsf{n}\not\in \varphi,\sigma$})
\end{align*}
\end{itemize}
\end{rem}
We now use our observations to derive some simple but useful validities.  
\begin{prop}
The following are $\mathsf{MLSR}$-provable validities: \begin{itemize}
\item[\emph{(i)}] $\langle!\varphi\rangle \alpha \rightarrow \varphi$ (announced formulas are always true);
\item[\emph{(ii)}] $\langle -(\varphi_{1}\vee\varphi_{2})\rangle\psi \leftrightarrow \big(\langle -\varphi_{1}\rangle\psi \vee \langle -\varphi_{2}\rangle\psi\big)$ (distributivity over disjunction, cf. Fact \ref{VDISTR});  
\item[\emph{(iii)}] $\mathsf{E}\alpha\leftrightarrow (\alpha\vee\langle-\alpha\rangle\top)$ (the removal modality captures quantifiers).
\end{itemize}
\end{prop}
\begin{proof}
(i) This follows since $\langle!\varphi\rangle \alpha \rightarrow \langle!\varphi\rangle\top$ is provable by principles of the minimal logic $\mathsf{K}$  for the  modality $\langle!\varphi\rangle$, while the $\mathsf{PAL}$ reduction axiom for the atom $\top$ gives   $\langle!\varphi\rangle\top \leftrightarrow \varphi$.

(ii) With the Basic Mix Rule in hand, it is straightforward to derive this non-trivial distribution law. We sketch the left-to-right direction, appealing to the Basic Mix Rule  with $\alpha= \varphi_{1}\vee\varphi_{2}$ and $\sigma=\del{\varphi_{1}}\psi \,\vee\del{\varphi_{2}}\psi$.
For $\mathsf{k}$ a fresh nominal, $\mathsf{E} (\mathsf{k} \land (\varphi_{1}\vee\varphi_{2})) \land \langle !\neg\mathsf{k}\rangle
\rangle\psi)$ provably implies $\del{\varphi_{1}}\psi \,\vee\del{\varphi_{2}}\psi$: this can be shown using the standard distribution of the $\mathsf{E}$ modality over disjunctions, after which the Mix Axiom gives the required result.

(iii) For the left-to-right direction, let $\mathsf{k}$ be a fresh nominal not appearing in $\alpha$. Note that, by the (H) Axiom, $\mathsf{E}(\mathsf{k}\wedge\alpha)\rightarrow \mathsf{U}(\neg\alpha\rightarrow\neg\mathsf{k})$ is derivable. Then, since 
\begin{align*}    
&\vdash (\mathsf{E}(\mathsf{k}\wedge\alpha) \wedge\neg\alpha)\rightarrow (\mathsf{E}(\mathsf{k}\wedge\alpha) \wedge\neg\mathsf{k}),\\
&\vdash (\mathsf{E}(\mathsf{k}\wedge\alpha) \wedge\neg\mathsf{k})\rightarrow (\mathsf{E}(\mathsf{k}\wedge\alpha) \wedge\bang{\neg\mathsf{k}}\top),\text{ and}\\
&\vdash (\mathsf{E}(\mathsf{k}\wedge\alpha) \wedge\bang{\neg\mathsf{k}}\top)\rightarrow \langle-\alpha\rangle\top \,\,\,(\text{by the Mix Axiom}),
\end{align*}
we have that $\vdash (\mathsf{E}(\mathsf{k}\wedge\alpha) \wedge\neg\alpha)\rightarrow \langle-\alpha\rangle\top$. The following instance of the Basic Paste Rule then gives us the desired conclusion: 
$$\infer[(\mathsf{k}\not\in \alpha)]{\mathsf{E}\alpha\rightarrow (\alpha\vee\langle-\alpha\rangle\top)}{\mathsf{E}(\mathsf{k}\wedge\alpha)\rightarrow(\alpha\vee\langle-\alpha\rangle\top)}$$
For the right-to-left direction, we have to show that $\langle-\alpha\rangle\top\rightarrow\mathsf{E}\alpha$. Let $\mathsf{k}$ be a fresh nominal not appearing in $\alpha$. Since $(\mathsf{E}(\mathsf{k}\wedge\alpha)\wedge \bang{\neg\mathsf{k}}\top)\rightarrow\mathsf{E}\alpha$ is clearly derivable, this follows from the  instance of the Basic Mix Rule displayed here:

\smallskip

$$\infer[(\mathsf{k}\not\in \alpha)]{ \langle-\alpha\rangle\top\rightarrow\mathsf{E}\alpha}{  (\mathsf{E}(\mathsf{k}\wedge\alpha)\wedge \bang{\neg\mathsf{k}}\top)\rightarrow\mathsf{E}\alpha}$$

\vspace{-2mm}

\end{proof}

Many of the formal proof routines illustrated in this section will be assumed without further explanation in the completeness proof of our next section. 

The above axiom system, though matching our later completeness proof,  may have some redundancies in its formulation. There is more power to the $\mathsf{PAL}$ reduction axioms than meets the eye, and the same is true of the Mix Rule. 

\begin{rem}
Consider the Truth Axiom, a modest, but useful principle:
$$\bang{\top}\varphi \leftrightarrow \varphi$$
In public announcement logic $\mathsf{PAL}$ with nominals and global modalities, the Truth Axiom is redundant, as all its instances are derivable. This can  be shown by a straightforward induction on the formula $\varphi$. The base cases for atoms (proposition letters, nominals and $\top$), as well as the inductive steps for negations, disjunctions, and the two existential modalities are immediate from the corresponding reduction axioms in  $\mathsf{PAL}$.

However, in the setting of $\mathsf{MLSR}$, we must also consider the inductive step for the removal modality. As it happens, one direction presents no difficulties. By the Stripped Mix Rule, to prove $\bang{\top}\langle-\alpha\rangle\psi \rightarrow \langle-\alpha\rangle\psi$, it suffices to derive, for some fresh nominal $\mathsf{k}$, the implication  $\bang{\top}(\mathsf{E}(\mathsf{k} \land \alpha) \land \bang{\neg \mathsf{k}}\psi) \rightarrow \langle-\alpha\rangle\psi$. And here, distributing the modality $\bang{\top}$ inside by appealing to the $\mathsf{PAL}$ axioms of $\mathsf{MLSR}$, and using the inductive hypothesis that $\bang{\top}\alpha \leftrightarrow \alpha$ is derivable already,  the antecedent is   provably equivalent to $\mathsf{E}(\mathsf{k} \land \alpha) \land \bang{\neg \mathsf{k}}\psi$, which implies $\langle-\alpha\rangle\psi$ by the Mix Axiom.

A similar analysis in the opposite direction would derive $\langle-\alpha\rangle\psi \rightarrow \bang{\top}\langle-\alpha\rangle\psi$ using the earlier Basic Mix Rule. However, showing the validity of that rule involved an appeal to the Truth Axiom, and it is not clear whether we can do without.
\end{rem}

We leave finding a more minimal and provably non-redundant presentation of $\mathsf{MLSR}$ as an open problem (see also the final point in \S\ref{Completness} about the need for the $\mathsf{PAL}$ component). Even so, as shown in this section, $\mathsf{MLSR}$ is quite a workable proof system, whose fine-structure deserves further exploration.

\section{Completeness}\label{Completness}
We now proceed to prove (strong) completeness of our deductive calculus.
\begin{thm}\label{completeness thm}
The system \emph{\textsf{MLSR}} is complete for validity in the given semantics.
\end{thm}

Soundness of the given axioms and rules follows from a straightforward inspection. The Henkin-style completeness proof follows standard modal and hybrid lines \citep{BRV:2001}, but there are some interesting new features that will be highlighted in what follows. We begin with a  preliminary definition toward a Lindenbaum Lemma. 

\begin{defn}[Named, Pasted, Mixed]\label{DEFMIX}
A set of \emph{\textsf{MLSR}}-formulas $\Gamma$ is
\begin{itemize}
    \item \emph{named} if it contains a nominal;
    \item $\Diamond$-\emph{pasted} if  $\mathsf{E}(\mathsf{n}\land\Diamond\varphi)\in\Gamma$ implies that there is some nominal $\mathsf{m}$ such that the formula $\mathsf{E}(\mathsf{n}\land\Diamond\mathsf{m})\wedge \mathsf{E}(\mathsf{m}\land\varphi)\in\Gamma$;
    \item $\mathsf{E}$-\emph{pasted} if  $\mathsf{E}(\mathsf{n}\land\mathsf{E}\varphi)\in\Gamma$ implies that there is some nominal $\mathsf{m}$ such that the formula $\mathsf{E}(\mathsf{n}\land\mathsf{E}\mathsf{m})\wedge \mathsf{E}(\mathsf{m}\land \varphi)\in\Gamma$;
\item \emph{mixed} if $\bang{\varphi}\langle-\alpha\rangle\psi\in\Gamma$ implies that there is some nominal $\mathsf{n}$ such that the formula $\bang{\varphi} \mathsf{E}(\mathsf{n}\wedge\alpha)\wedge\bang{\varphi}\langle!\neg\mathsf{n}\rangle\psi\in\Gamma$;
\item $\mathsf{E}$\emph{-mixed} if, whenever $\mathsf{E}(\mathsf{n}\wedge \bang{\varphi}\del{\alpha}\psi)\in \Gamma$, then there is some nominal  $\mathsf{k}$ such that $\mathsf{E}\big(\mathsf{n}\wedge\bang{\varphi} (\mathsf{E}(\mathsf{k}\wedge\alpha)\wedge\bang{\neg\mathsf{k}}\psi)\big)\in\Gamma$.
\end{itemize}
\end{defn}

A set $\Gamma$ of \textsf{MLSR}-formulas will be said to be  \emph{pasted} if it is both $\Diamond$-pasted and $\mathsf{E}$-pasted. The technical reason for having  two mixing principles instead of one  will become clear later on. But it may be noted here already that if a deductively closed set $\Gamma$ contains some nominal $\mathsf{n}$ naming it, then $\mathsf{E}$-mixing implies plain mixing. For, in that case, as was shown in Remark \ref{weak hybrid}, modulo $\Gamma$, any formula $\psi$ will be provably equivalent to $\mathsf{E}(\mathsf{n} \land \psi$).

\begin{rem}As will be seen below, the Mix Rule of $\mathsf{MLSR}$ supports the preceding `mixing': i.e., witnessing the removal modality by introducing a new nominal for the point to be removed. As stated, the rule does this  only under one-step update modalities $\langle !\varphi\rangle$. But this implies the Mix Rule for arbitrary finite sequences of updates. First, the special case of $\langle !\top\rangle\psi$ gives the case of single formulas $\psi$, as the two are equivalent in \emph{\textsf{MLSR}}. But also, longer sequences of updates are covered, as is easy to see using the  \emph{\textsf{PAL}} axiom $\bang{\varphi}\bang{\psi}\alpha \leftrightarrow \bang{(\varphi\wedge [!\varphi]\psi)}\alpha$  compressing two nested update modalities to a single one. 

It follows that if a deductively closed set $\Gamma$ is mixed, then it also  witnesses \emph{sequences} of announcement modalities $\bang{\overline{\varphi}}$, where $\bang{\overline{\varphi}}:=\bang{\varphi_{1}}...\bang{\varphi_{k}}$ for a sequence of formulas $\overline{\varphi} = (\varphi_{1},\hdots,\varphi_{k})$. For instance, with simple mixing: if $\bang{\overline{\varphi}}\langle-\alpha\rangle\psi\in\Gamma$, then there is a nominal $\mathsf{n}$ such that $\bang{\overline{\varphi}} \mathsf{E}(\mathsf{n}\wedge\alpha)\wedge\bang{\overline{\varphi}}\langle!\neg\mathsf{n}\rangle\psi\in\Gamma$.\footnote{For a concrete case of how this works, suppose that $\bang{\varphi_1}\bang{\varphi_2}\del{\alpha}\psi\in\Gamma$. Using the $\mathsf{PAL}$ iteration axiom $\bang{\varphi_1}\bang{\varphi_2}\vartheta \leftrightarrow \bang{(\varphi_1\wedge [!\varphi_1]\varphi_2)}\vartheta$, we get  $\bang{(\varphi_1\wedge [!\varphi_1]\varphi_2)}\del{\alpha}\psi\in\Gamma$.
Since $\Gamma$ is mixed,  there is then a nominal \textsf{n} such that $ \bang{(\varphi_1\wedge [!\varphi_1]\varphi_2)}\textsf{E}(\textsf{n}\wedge\alpha)\wedge \bang{(\varphi_1\wedge [!\varphi_1]\varphi_2)}\bang{\neg\textsf{n}}\psi\in\Gamma$. But then, using the $\mathsf{PAL}$ iteration axiom once more, it follows that $\bang{\varphi_1}\bang{\varphi_2}\textsf{E}(\textsf{n}\wedge\alpha)\wedge \bang{\varphi_1}\bang{\varphi_2}\bang{\neg\textsf{n}}\psi\in\Gamma$.} 
\end{rem}

\begin{lem}[Lindenbaum Lemma]
Every \emph{\textsf{MLSR}}-consistent set of formulas can be extended to an \emph{\textsf{MLSR}} maximal consistent set that is named, pasted, as well as mixed in both senses.
\end{lem}
\begin{proof}
Naming and pasting work in exactly the same way as in the  completeness proof for the basic hybrid logic. As for mixing, given the above observation, we only consider the case of  $\mathsf{E}$-mixing.
We have to ensure that, throughout the inductive construction, whenever we consistently add a formula of the form $\mathsf{E}(\mathsf{n} \land \bang{\varphi}\langle-\alpha\rangle\psi)$ to a consistent, named set of formulas $\Sigma$,  the formula $\mathsf{E}(\mathsf{n}\land \bang{\varphi}(\mathsf{E}(\mathsf{k} \land \alpha) \land \langle !\neg \mathsf{k} \rangle \psi))$ is also added to $\Sigma$\textemdash where $\mathsf{k}$ is the first nominal in the  enumeration of nominals used in our construction that occurs in neither $\Sigma$ nor $\bang{\varphi}\langle-\alpha\rangle\psi$. Crucially, for such a $\mathsf{k}$,  the set $\Sigma \cup \{\mathsf{E}(\mathsf{n}\land \bang{\varphi}(\mathsf{E}(\mathsf{k} \land \alpha) \land \langle !\neg \mathsf{k} \rangle \psi))\}$ is consistent, given that $\Sigma$ is consistent.  For if not, then for some conjunction $\sigma$ of formulas from $\Sigma$, the implication $\mathsf{E}(\mathsf{n}\land \bang{\varphi}(\mathsf{E}(\mathsf{k} \land \alpha) \land \langle !\neg \mathsf{k} \rangle \psi)) \rightarrow \neg \sigma $ would be provable. But then, by the Mix Rule,  the implication $\mathsf{E}(\mathsf{n} \land \bang{\varphi}\langle-\alpha\rangle\psi) \rightarrow \neg \sigma$ is provable from $\Sigma$, contradicting our initial assumption that $\Sigma\cup\{\mathsf{E}(\mathsf{n}\wedge\bang{\varphi}\langle-\alpha\rangle\psi)\}$ is consistent.
\end{proof}

For the remainder of this proof, fix a maximal consistent set $\Gamma$ of  \textsf{MLSR}-formulas (an   $\textsf{MLSR}$-MCS, for short) that is named, pasted, and mixed in all the senses of Definition \ref{DEFMIX}. Next, for all nominals $\mathsf{n}$ with $\mathsf{E}\mathsf{n} \in\Gamma$, define the set $\Delta_{\mathsf{n}}:=\{\varphi\in \mathsf{MLSR}\,|\, \mathsf{E}(\mathsf{n}\land \varphi)\in \Gamma \}$.  Let $$\mathcal{W}= \{\Gamma\}\cup \{\Delta_{\mathsf{n}}\,|\,\mathsf{n}\in\mathsf{NOM},\,\mathsf{E}\mathsf{n}\in\Gamma \}.$$ 

\noindent Over this universe, accessibility relations are defined as follows:
\begin{alignat*}{2}
 \mathcal{R}_{\Diamond}(\Delta_\mathsf{n}, \Delta_\mathsf{m}) \, &\text { iff} &\mathsf{E}(\mathsf{n} \land \Diamond\mathsf{m}) \in \Gamma \\
\mathcal{R}_{\mathsf{E}}(\Delta_\mathsf{n}, \Delta_\mathsf{m}) \, &\text { iff }  &\mathsf{E}(\mathsf{n} \land \mathsf{E}\mathsf{m}) \in \Gamma.
\end{alignat*}

\vspace{0.5mm}

In this definition, the set $\Gamma$ is thought of as containing all information about the whole universe  $\mathcal{W}$. This  includes information about $\Gamma$ itself, since, by an earlier observation, $\Gamma = \Delta_\mathsf{n}$ for any nominal $\mathsf{n}\in \Gamma$, and such nominals exist since $\Gamma$ is named by our Lindenbaum construction. One could continue the completeness argument in this style, but in what follows we consider all sets introduced here on a par, as `worlds' or `states' in a modal model, for which we will use the standard notation $w, v, ...$\footnote{The above accessibility relations could also be defined equivalently in a standard modal manner: for any $w,v \in \mathcal{W}$, we have  $\mathcal{R}_{\Diamond}(w, v)$ if and only if $\text{ for all }\varphi,\, \text{ if } \varphi\in v \text{ then } \Diamond\varphi\in w$, and similarly for $\mathcal{R}_{\mathsf{E}}$.}

\smallskip

We now define an initial structure toward finding a model for our consistent set.

\begin{defn}[Upper Henkin Model]
The \emph{upper Henkin model} $\mathcal{M}$ generated by $\Gamma$ is defined as the structure $(W, R_{\Diamond}, R_{\mathsf{E}}, V)$, where 
\begin{itemize}
    \item $W:= [\Gamma]_{\mathcal{R}_{\mathsf{E}}}$, the equivalence class of $\Gamma$ in $\mathcal{W}$ under $\mathcal{R}_{\mathsf{E}}$;
    \item the relations $R_{\Diamond}$ and $R_{\mathsf{E}}$ are, respectively, $\mathcal{R}_{\Diamond}$ and $\mathcal{R}_{\mathsf{E}}$ restricted to $W$;
    \item the valuation $V$ is given by $V(p) = \{w\in W\,|\,p\in w\}$ for all proposition letters $p$ and, for all nominals $\mathsf{n}, 
    V(\mathsf{n}) = 
   \{\Delta_\mathsf{n}\} \, \text{if }\, \mathsf{En}\in\Gamma,\text{and } V(\mathsf{n}) = \varnothing \,\, \text{otherwise.}
    $
\end{itemize}
\end{defn}
Setting the domain to be the equivalence class of $\Gamma$ under $\mathcal{R}_{\mathsf{E}}$ ensures that $R_{\mathsf{E}}$ is the universal relation in the model. It is also easy to show that the valuation is well-defined for nominals. This construction has some important properties, listed in the next result.

\begin{lem}[Existence Lemma]\label{EXLEM}
Let $\Gamma$ be a named, pasted and $\mathsf{E}$-mixed \emph{\textsf{MLSR}}-MCS and $\mathcal{M}= (W, R_\Diamond, R_\mathsf{E}, V)$ the upper Henkin model yielded by $\Gamma$. 
\begin{itemize} 
\item All sets $\Delta_\mathsf{n}$ are \emph{\textsf{MLSR}}-MCSs;
\item If $u\in W$ and $\Diamond\varphi\in u$, then there is an object $v\in W$ such that $R_\Diamond uv$ and $\varphi\in v$. 

\item If $u\in W$ and  $\mathsf{E}\varphi\in u$, then there is some $v\in W$ such that $R_\mathsf{E}uv$ and $\varphi \in v$.
\item All sets $\Delta_\mathsf{n}$ are mixed in the first sense listed in Definition \ref{DEFMIX}.
\end{itemize}
\end{lem}

The proof of all these assertions is by reference to the Pasting and $\mathsf{E}$-Mixing properties of the original set $\Gamma$, using principles available in $\mathsf{MLSR}$ that were identified earlier.

\medskip

Next, we define a family of derived structures which  capture the effects of finite sequences of updates on the  upper Henkin model $\mathcal{M}$. Recall that, given a sequence $\overline{\varphi} = (\varphi_{1},\hdots,\varphi_{k})$, the notation $\bang{\overline{\varphi}}$ stands for $\bang{\varphi_{1}}...\bang{\varphi_{k}}$. 

\begin{defn}[Derived Henkin Model]
 For each finite sequence of $\mathsf{MLSR}$-formulas $\overline{\varphi}= 
(\varphi_{1},\hdots,\varphi_{k})$,  the \emph{derived Henkin model} $\mathcal{M}:\overline{\varphi}$ is the structure $(W^{\overline{\varphi}}, R_\Diamond^{\overline{\varphi}}, R^{\overline{\varphi}}_{\mathsf{E}}, V^{\overline{\varphi}})$, where
 \begin{itemize}
     \item $W_{\overline{\varphi}}:=\{(w,\overline{\varphi})\,|\, w\in W \text{ and }\bang {\varphi_{1}}\hdots\bang {\varphi_{k}}\top \in w \}$;
    \item for $(w, \overline{\varphi}), (v, \overline{\varphi}) \in W_{\overline{\varphi}}$, $R_\Diamond^{\overline{\varphi}}\big((w,\overline{\varphi}) , (v,\overline{\varphi})\big)$ (resp., $R^{\overline{\varphi}}_{\mathsf{E}}\big((w,\overline{\varphi}) , (v,\overline{\varphi})\big)$) if and only if $R_{\Diamond}wv$ (resp., $R_{\mathsf{E}}wv$) in the upper Henkin model $\mathcal{M}$; 
    \item $V^{\overline{\varphi}}(p):= \{(w,\overline{\varphi})\,|\, p\in w\} $ for $p\in\mathsf{PROP}$ and $V^{\overline{\varphi}}(\mathsf{n}):= \{(w,\overline{\varphi})\,|\, \mathsf{n}\in w\} $ for $\mathsf{n}\in\mathsf{NOM}$.
 \end{itemize}
\end{defn}

Points in the derived Henkin model $\mathcal{M}:\overline{\varphi}$ are  sequences $(w, \varphi_{1},\hdots,\varphi_{k})$, where $w$ is a MCS in the upper Henkin model that contains the \emph{pre-condition formula}\footnote{From another perspective, worlds in derived Henkin models are like the finite update histories in temporal `protocol models' for \textsf{PAL} \citep{BGHP:2009}.} $$\mathsf{pre}(\overline{\varphi}) = \bang{\overline{\varphi}}\top = \langle !\varphi_{1} \rangle \hdots\langle !\varphi_{k}\rangle \top $$  The accessibility relations stay as they were for the initial points of the sequences in the upper Henkin model. Likewise, the valuation for proposition letters at each sequence stays the same as that for its initial point in the upper Henkin model. 
 
In  derived Henkin models, all points are still named by nominals, but some nominals may fail to denote. This explains the modified hybrid base logic for \textsf{MLSR} (Remark \ref{weak hybrid}).
 
Now, to each point $(w, \varphi_{1},\hdots,\varphi_{k})$ in the derived Henkin model $\mathcal{M}:\overline{\varphi}$, we associate the following set of formulas 
$$\Phi(\mathcal{M}, \overline{\varphi},w):=\{\alpha\,|\, \bang{\varphi_{1}}\hdots\bang{\varphi_{k}}\alpha\in w \}.$$

\noindent These sets record what the upper Henkin model `claims' is true after the update with $\overline{\varphi}$. Our task is to analyze how this matches up with truth in the updated models.

To do so, we first note some useful theorems of \textsf{MLSR} concerning the effects of finite sequences of successive updates. The first of these computes preconditions explicitly: 

\vspace{-4mm}

\begin{align*}
\vdash_{\textsf{MLSR}} \langle!\varphi_1\rangle\hdots\langle!\varphi_k\rangle\top &\leftrightarrow\varphi_1\wedge \langle!\varphi_1\rangle\varphi_2 \wedge\langle!\varphi_1\rangle\langle!\varphi_2\rangle\varphi_3\wedge\hdots\wedge \langle!\varphi_1\rangle\hdots\langle!\varphi_{k-1}\rangle\varphi_k\\
&\leftrightarrow \bigwedge_{i=1}^k\langle!\varphi_1\rangle\hdots\langle!\varphi_{i-1}\rangle\varphi_i  \label{eq:r1} \tag{\textsf{R1}}
\end{align*}
The derivation of \eqref{eq:r1} and the following principles of \textsf{MLSR} are obtained by straightforward iteration of the principles of public announcement logic  \textsf{PAL} \citep{JB:2011}. 

Our second observation analyses when atomic formulas are true after iterated updates:

\begin{equation} \label{eq:r2} \tag{\textsf{R2}}
\vdash_{\textsf{MLSR}} \langle!\overline{\varphi}\rangle p \leftrightarrow  (\mathsf{pre}(\overline{\varphi})\wedge p)
\end{equation}  

\noindent The preceding theorem also applies to nominals. A similar pattern occurs with negations:\footnote{For a concrete illustration, the following chain of equivalences is provable:
$\langle !\varphi_{1} \rangle \langle !\varphi_{2}\rangle \neg \psi \, \leftrightarrow \, \langle !\varphi_{1}\rangle(\varphi_{2} \land \neg \langle !\varphi_{2}\rangle\psi) \, \leftrightarrow \, (\langle !\varphi_{1}\rangle \varphi_{2} \land \langle !\varphi_{1}\rangle\neg \langle!\varphi_{2}\rangle\psi) \, \leftrightarrow \,
(\langle !\varphi_{1}\rangle\varphi_{2} \land \neg \langle !\varphi_{1}\rangle \langle!\varphi_{2}\rangle\psi$).} 
\begin{equation}
\vdash_{\textsf{MLSR}} \langle!\overline{\varphi}\rangle \neg\alpha \leftrightarrow  (\mathsf{pre}(\overline{\varphi})\wedge  \neg\langle!\overline{\varphi}\rangle\alpha)\label{eq:r3} \tag{\textsf{R3}}
\end{equation}

\noindent For conjunctions, it is easy to see that 
\begin{equation} \label{eq:r4} \tag{\textsf{R4}}
\vdash_{\textsf{MLSR}} \langle!\overline{\varphi}\rangle (\alpha \wedge \beta) \leftrightarrow  (\langle!\overline{\varphi}\rangle\alpha \wedge \langle!\overline{\varphi}\rangle\beta)
\end{equation}
\noindent Finally, consider the diamond modality. Here we have:

\begin{equation}  \label{eq:r5} \tag{\textsf{R5}}
\vdash_{\textsf{MLSR}} \langle!\overline{\varphi}\rangle\Diamond\alpha \leftrightarrow  (\mathsf{pre}(\overline{\varphi})\wedge  \Diamond\langle!\overline{\varphi}\rangle\alpha)
  \end{equation}

 \noindent A similar principle holds for global existential modalities of the form \textsf{E}$    \alpha$.\footnote{All these facts unpack finite sequences of \textsf{PAL}-update modalities. But they can also be understood in terms of one-step modalities using the \textsf{PAL} axiom for compressing two iterated updates into one.}

\medskip
 
Intuitively, the models $\mathcal{M}:\overline{\varphi}$ are meant to be  isomorphic to submodels of $\mathcal{M}$ after the sequence of consecutive semantic updates $ \overline{\varphi}$, but the precise sense in which this is true will become clear in the following key property of the  construction of initial and derived Henkin models.\footnote{While not essential for what follows, the following fact  further clarifies the structure of derived Henkin models. Given $(w,\overline{\varphi})$ and $(v,\overline{\varphi})$ in $W_{\overline{\varphi}}$, the following assertions are equivalent: (a) $R_\Diamond^{\overline{\varphi}}\big((w :\overline{\varphi}), (v :\overline{\varphi})\big)$ as defined earlier,
     (b) for all formulas  $\alpha$, if $\alpha\in \Phi(\mathcal{M},\overline{\varphi}, v), \text{then } \Diamond\alpha\in \Phi(\mathcal{M},\overline{\varphi}, w)$. We omit the proof here.}
     
Now comes the main point of our construction so far.

\begin{lem}[Truth Lemma]
For all formulas $\psi$, finite sequences $\overline{\varphi}$ and points $w$,

\vspace{-1mm}

 $$ \mathcal{M}:\overline{\varphi}, (w, \overline{\varphi}) \models \psi \,\,\,  \text{if and only if } \,\,\, \psi \in \Phi(\mathcal{M}, \overline{\varphi}, w) $$
\end{lem}

\begin{proof}
The proof is by induction on the formulas $\psi$. For convenience, when the context is clear, we write $w$ instead of $(w, \overline{\varphi})$, reflecting the fact that derived Henkin models represent submodels arising from iterated \textsf{PAL} updates. Also note that, by the earlier definitions, the existence of the state $(w, \overline{\varphi})$ in $\mathcal{M}:\overline{\varphi}$ assumes that the precondition  $\mathsf{pre}(\overline{\varphi})$ of the relevant update sequence  belongs to $w$. This fact will be used repeatedly in what follows.

\medskip

\noindent(a) For the equivalence of truth and membership for atoms $p$, it suffices to observe that  $ p \in \Phi(\mathcal{M}, \overline{\varphi}, w)$ iff $\langle !\overline{\varphi} \rangle p \in w $  iff (by the above-noted fact \eqref{eq:r2})  $ \mathsf{pre}(\overline{\varphi}) \land p \in w$. But then, by the definition of the valuation in derived Henkin models, this means that $p$ is true at the initial  $w$ and all its descendants under update. The same argument applies to nominals.

\medskip

\noindent(b) For negations, we have $\neg \psi \in \Phi(\mathcal{M}, \overline{\varphi}, w)$ iff 
$\langle !\overline{\varphi} \rangle \neg \psi \in w $, which, by \eqref{eq:r3}, is provably equivalent to $\mathsf{pre}(\overline{\varphi}) \in w$ and $\neg \langle !\overline{\varphi} \rangle \psi \in w $. Given that $ \mathsf{pre}(\overline{\varphi}) \in w$, we have that 
$\langle !\overline{\varphi} \rangle \neg \psi \in w $ iff $\langle !\overline{\varphi} \rangle \psi \notin w $. The latter statement is equivalent to $\psi\not\in \Phi(\mathcal{M},\overline{\varphi}, w)$, which, by the inductive hypothesis for $\psi$, holds if and only if $\mathcal{M}:\overline{\varphi},w\models \neg\psi$.

\medskip
\noindent(c) The inductive step for conjunctions $\psi_{1} \wedge \psi_{2}$ is straightforward, using the above distribution principle \eqref{eq:r4} of $\langle !\overline{\varphi} \rangle$  modalities over conjunctions, as well as the fact that maximally consistent sets decompose conjunctions into components.

\medskip

\noindent(d) Next, we consider the  basic modality $\Diamond\psi$, relying on the above principle \eqref{eq:r5}. 

\begin{itemize}
    \item If $\mathcal{M}:\overline{\varphi}, w \models \Diamond\psi$, then, for some $v$ with $R_\Diamond^{\overline{\varphi}}wv$, $\mathcal{M}:\overline{\varphi}, v \models \psi$. So, by the inductive hypothesis, $\psi\in \Phi(\mathcal{M}, \overline{\varphi}, v)$. Hence, in the upper Henkin model $\mathcal{M}$, we have that $\bang{\overline{\varphi}}\psi\in v$ and also $R_\Diamond wv$. This entails that, in $\mathcal{M}$, $\Diamond \langle ! \overline{\varphi}\rangle \psi\in w$. Now, using \eqref{eq:r5} and the fact that  $\mathsf{pre}(\overline{\varphi})  \in w$,  $\bang{\overline{\varphi}}\Diamond\psi\in w$. By the definition of $\Phi$, then $\Diamond\psi\in \Phi(\mathcal{M}, \overline{\varphi}, w)$.
    \item Conversely: $\Diamond\psi\in \Phi(\mathcal{M}, \overline{\varphi}, w)$ entails that 
    $\bang{\overline{\varphi}}\Diamond\psi\in w$ in the upper Henkin model $\mathcal{M}$. From \eqref{eq:r5} we obtain $\Diamond \langle ! \overline{\varphi}\rangle \psi\in w$. Therefore, by the Existence Lemma \ref{EXLEM}, there is some $v\in \mathcal{M}$ with $R_\Diamond wv$ and $\bang{\overline{\varphi}}\psi\in v$. Now obviously $\vdash_{\mathsf{MLSR}}\bang{\overline{\varphi}}\psi\rightarrow \bang{\overline{\varphi}}\top =  \mathsf{pre}(\overline{\varphi})$, and so it follows that $\mathsf{pre}(\overline{\varphi})\in v$. Hence, $v$ is in the  derived Henkin model under consideration here.
	So, we have that $\psi\in \Phi(\mathcal{M}, \overline{\varphi}, v)$ and, by the inductive hypothesis, we get $\mathcal{M}:\overline{\varphi}, v\models \psi$. Since $R_\Diamond wv$, we have $R_\Diamond^{\overline{\varphi}}wv$, and so $\mathcal{M}:\overline{\varphi}, w \models \Diamond\psi$.
	\end{itemize}

\noindent(e) The reasoning for the existential modality $\mathsf{E}\psi$ is just like the preceding argument.
    
    \medskip
\noindent(f) The analysis for \textsf{PAL} modalities $\bang{\alpha}\psi$ proceeds as follows. 
\medskip

\noindent First note that, by the inductive hypothesis, the truth of $\alpha$ at any point $(v, \overline{\varphi})$ is equivalent to $\alpha$ belonging to $ \Phi(\mathcal{M}, \overline{\varphi}, v)$. Hence, restricting the model $ \mathcal{M}:\overline{\varphi}$ to $ (\mathcal{M}:\overline{\varphi})|\alpha$ in the usual semantic sense yields exactly the derived Henkin model $ \mathcal{M}:(\overline{\varphi}^{\frown}\alpha)$.\footnote{Thanks to the inductive hypothesis, the model $ (\mathcal{M}:\overline{\varphi})|\alpha$ contains exactly the states $w$ for which $\bang{\overline{\varphi}}\alpha\in w$. But by its definition, the derived Henkin model $\mathcal{M}:(\overline{\varphi}^{\frown}\alpha)$ is restricted to exactly the states $w$ such that $\mathsf{pre}(\overline{\varphi}^{\frown}\alpha)\in w$, which is equivalent to $\bang{\overline{\varphi}}\alpha\in w$.}\\
We then have the following equivalences:
\begin{align*}
\mathcal{M}:\overline{\varphi}, w\models \bang{\alpha}\psi \quad &\text{ iff }  \mathcal{M}:\overline{\varphi}, w\models\alpha \text{ and } (\mathcal{M}:\overline{\varphi}) |\alpha, w\models \psi & \\
&\text{ iff } \mathcal{M}:(\overline{\varphi}^{\frown}\alpha), w\models \psi &\\
&\text{ iff } \psi\in\Phi(\mathcal{M}, \overline{\varphi}^{\frown}\alpha , w) & \text{(by the inductive hypothesis)}\\
& \text{ iff } \bang{\alpha}\psi\in \Phi(\mathcal{M},\overline{\varphi}, w) & \text{(since $\bang{\overline{\varphi}^{\frown}\alpha}\psi = \bang{\overline{\varphi}}\bang{\alpha}\psi$) }
\end{align*}

\noindent(g) Finally, the \textsf{MLSR} deletion modality $\del{\alpha}\psi$ is analyzed as follows. 
\begin{itemize}
\item If $\mathcal{M}:\overline{\varphi}, w \models \del{\alpha}\psi$, then, by the truth definition, for some $v\neq w$, (i) $\mathcal{M}:\overline{\varphi}, v \models \alpha$ and (ii) $(\mathcal{M}:\overline{\varphi})~-~\{v\}, w\models~ \psi$. Next, since all states in the upper Henkin model are named (see Definition \ref{DEFMIX}), there exists a nominal   $\mathsf{n}$ denoting $v$, and this nominal is still available for denoting $v$'s descendant in the derived Henkin model $\mathcal{M}:\overline{\varphi}$. 

\smallskip
First consider conjunct (i). By the inductive hypothesis, $\alpha\in\Phi(\mathcal{M}, \overline{\varphi}, v)$, while also, for our nominal $\mathsf{n}$, we have  $\mathsf{n}\in\Phi(\mathcal{M}, \overline{\varphi}, v)$. Therefore, in the upper Henkin model $\mathcal{M}$, $ \bang{\overline{\varphi}} (\mathsf{n}\wedge \alpha)\in v $. Now, the relation $R_{\mathsf{E}}$ is universal in the upper Henkin model, and so, in particular, $R_{\mathsf{E}}wv$, which entails that $\mathsf{E} \bang{\overline{\varphi}} (\mathsf{n}\wedge \alpha)\in w$. Next, by our earlier observations about provable generalized reduction axioms in \textsf{MLSR}:
\begin{equation}  \label{eq:RE} \tag{$\dagger$}
  \vdash_{\mathsf{MLSR}} \bang{\overline{\varphi}}\mathsf{E}(\mathsf{n}\wedge \alpha)\leftrightarrow \big(\mathsf{pre}(\overline{\varphi})\wedge \mathsf{E} \bang{\overline{\varphi}} (\mathsf{n}\wedge \alpha) \big)
  \end{equation}
Together with the fact that  $\mathsf{pre}(\overline{\varphi}) \in w$, it follows that  $\bang{\overline{\varphi}}\mathsf{E}(\mathsf{n}\wedge \alpha)\in w$.  

\smallskip
Next, consider conjunct (ii) in our initial assumption. It is easy to see that the model $(\mathcal{M}:\overline{\varphi})~-~\{v\}$ equals 
the updated model $(\mathcal{M}:\overline{\varphi})| \neg\mathsf{n}$, using the fact that each nominal belongs to at most one world in $\mathcal{M}$. Thus (ii) implies that $(\mathcal{M}:\overline{\varphi})| \neg\mathsf{n}, w \models \psi$. Moreover, it can be seen that the latter model in turn equals 
$\mathcal{M}:~(\overline{\varphi} ^ \frown \neg\mathsf{n})$.\footnote{Observe that $w\in\mathcal{M}:(\overline{\varphi} ^ \frown \neg\mathsf{n})$ iff $\mathsf{pre}(\overline{\varphi}^{\frown}\neg\mathsf{n}) \in w$ iff $\bang{\overline{\varphi}}\neg\mathsf{n}\in w$. Now, as observed earlier, $\vdash_{\mathsf{MLSR}}\bang{\overline{\varphi}}\neg\mathsf{n} \leftrightarrow \mathsf{pre}(\overline{\varphi}) \wedge \neg \bang{\overline{\varphi}}\mathsf{n} $, so the last statement is equivalent to $\mathsf{pre}(\overline{\varphi})\in w$ and $\bang{\overline{\varphi}}\mathsf{n}\not\in w$, which holds exactly when  $\mathsf{n}\not\in w \in \mathcal{M}:\overline{\varphi}$, which means that $w \in (\mathcal{M}:\overline{\varphi})| \neg\mathsf{n}$ by the truth definition.} Therefore, we also have $\mathcal{M}:~(\overline{\varphi} ^ \frown \neg\mathsf{n}), w\models\psi$. But then, by the inductive hypothesis, we have that $\psi\in \Phi(\mathcal{M}, \overline{\varphi}^{\frown}\neg\mathsf{n}, w)$, i.e., $\bang{\overline{\varphi}}\langle !\neg \mathsf{n} \rangle \psi \in w $.

It now remains to apply the Mix Axiom. As stated in the definition of the system \textsf{MLSR}, this says that the conjunction  $\mathsf{E}(\mathsf{n}\wedge \alpha)\wedge \bang{\neg\mathsf{n}}\psi$ implies 
 $\del{\alpha}\psi$. Now in the upper Henkin model, we only have the antecedent for this in the set $w$  under the prefix $\langle !\overline{\varphi}\rangle$.  But then the modalized conclusion $\langle !\overline{\varphi}\rangle \del{\alpha}\psi$  is derivable using the \textsf{K} axioms for the $\bang{\varphi}$ modalities, and hence it, too, is in $w$. In other words, $ \del{\alpha}\psi \in \Phi(\mathcal{M}, \overline{\varphi}, w)$.

\item Next suppose that $\del{\alpha}\psi \in \Phi(\mathcal{M}, \overline{\varphi},w )$: i.e., $\bang{\overline{\varphi}}\del{\alpha}\psi\in w$. Since $w$ is mixed by the Existence Lemma \ref{EXLEM}, there exists some nominal $\mathsf{n}$ such that (i) $\bang{\overline{\varphi}} \mathsf{E}(\mathsf{n}\wedge \alpha) \in w$, and (ii) $ \bang{\overline{\varphi}}\bang{\neg\mathsf{n}}\psi\in w$. Using the earlier equivalence (\ref{eq:RE}) once more, from (i), we get $\mathsf{E} \bang{\overline{\varphi}} (\mathsf{n}\wedge \alpha)\in w$. By the Existence Lemma \ref{EXLEM} once more, this means that there is some $v\in \mathcal{M}$ with $\bang{\overline{\varphi}} (\mathsf{n}\wedge \alpha)\in v$. So, $\mathsf{n} \wedge \alpha \in \Phi(\mathcal{M}, \overline{\varphi},v )$. By the inductive hypothesis, this entails that $\mathcal{M}:\overline{\varphi}, v\models \alpha$.  

\smallskip

Next, turning to (ii), using the straightforward observation that  $\vdash_{\mathsf{MLSR}} \bang{\overline{\varphi}}\neg\mathsf{n} \rightarrow  \neg \mathsf{n} $, we get $ \mathsf{n}\not\in w$, while $\bang{\overline{\varphi}} (\mathsf{n}\wedge \alpha)\in v$ entails $\mathsf{n}\in v$, so $w\neq v$. 

\smallskip

Lastly, $ \bang{\overline{\varphi}}\bang{\neg\mathsf{n}}\psi\in w$ means $\bang{\neg\mathsf{n}}\psi\in \Phi(\mathcal{M}, \overline{\varphi},w )$. This is equivalent to $\psi\in  \Phi(\mathcal{M}, \overline{\varphi}^{\frown}\neg \mathsf{n},w )$, which, by the inductive hypothesis, yields that $\mathcal{M}:\overline{\varphi}^{\frown}\neg \mathsf{n}, w\models \psi$. Next, as already noted in the argument for the converse direction, $\mathcal{M}:(\overline{\varphi}^{\frown}\neg \mathsf{n}) = (\mathcal{M}:\overline{\varphi})|\neg\mathsf{n}$, and hence we get $(\mathcal{M}:\overline{\varphi})|\neg\mathsf{n}, w\models\psi$. Moreover, we had  $(\mathcal{M}:\overline{\varphi})|\neg\mathsf{n}= (\mathcal{M}:\overline{\varphi})-\{v\}$, and so we also have $(\mathcal{M}:\overline{\varphi})-\{v\}, w\models \psi$.

\smallskip

Taking these three facts together, it can be concluded that $\mathcal{M}:\overline{\varphi}, w\models \del{\alpha}\psi$.
\end{itemize}
This concludes the proof of the Truth Lemma, and thus, the consistent set $\Gamma$ at the start of the construction has a model. This establishes the  completeness of the system \textsf{MLSR}. 
\end{proof}

\begin{rem}
One way of understanding the mechanics of this modal completeness proof is doing a parallel standard Henkin-style completeness proof for a first-order language with explicit operations of quantification without replacement and definable relativization.
\end{rem} 

Finally, a natural question is if our expanded language with nominals and \textsf{PAL} modalities is really needed. We leave the existence of a `pure' axiomatization of \textsf{MLSR} open here.\footnote{In ongoing follow-up work, Johan van Benthem, Li Lei, Chenwei Shi, and Haoxuan Yin at Tsinghua University have used the techniques presented here to axiomatize hybrid sabotage modal logic in several semantics, and to derive further results such as interpolation theorems. One feature of their approach suggests an alternative perspective on our completeness proof. The crucial use of public announcements in the above is in the special form $\langle! \neg \mathsf{n} \rangle \varphi$. This is equivalent to the hybrid \textsf{MLSR} formula $(\neg\mathsf{E}\mathsf{n}\wedge \varphi) \vee \langle - \mathsf{n}\rangle \varphi$. This suggests a pure axiomatization where the minimal \textsf{PAL} modalities become convenient suggestive notation. The full system \textsf{MLSR} merges definable uniform and arbitrary stepwise removal, thus describing quantification with and without replacement combined with unrestricted relativization.}

\section{Complexity and undecidability}\label{Complexity and undecidability}

Having analyzed expressive power and axiomatization, we now turn to matters of computational complexity for the core notions of our system $\mathsf{MLSR}$ as defined in \S\ref{Basics of expressive power}.  

\subsection{Model checking}

We begin by showing that model checking for $\mathsf{MLSR}$ is PSPACE-complete. We do so by providing a reduction from the quantified Boolean formula problem (QBF) \citep{SM:1973}, in the style of \cite{Roh:2005} and \cite{LR:2003}.

\begin{thm}\label{Model checking}
Model checking for $\mathsf{MLSR}$ is \emph{PSPACE}-complete.
\end{thm}

\begin{proof}
An upper bound is established as follows. The translation into first-order logic given earlier (Fact \ref{translation}) only has a polynomial size increase, and it is known that model checking for first-order logic is in PSPACE.

The lower bound is demonstrated by a reduction from QBF into model checking for \textsf{MLSR}. Take any QBF formula $\varphi$: that is, a formula of the form 
$$
Q_{1}x_{1}\,...\,Q_{n}x_{n} \bigwedge_{1\leq i\leq k} C_{i},
$$
where $Q_{j}\in\{\exists, \forall\}$, and each $C_{i}$ is a disjunction of literals $\pm x_{j}$ (here, without loss of generality, we can assume that the quantifiers alternate between $\exists$ and $\forall$).

Given such a formula $\varphi$, we construct a finite pointed model $(\mathcal{M}_{\varphi}, s)$ and an \textsf{MLSR} formula $\gamma_{\varphi}$ such that $\varphi$ is true if and only if $(\mathcal{M}_{\varphi}, s)\models\gamma_{\varphi}$. The construction will ensure that the model $\mathcal{M}_{\varphi}$ and the formula $\gamma_{\varphi}$ both have a size that grows linearly in the number of quantifiers and clauses of $\varphi$, which gives the desired reduction from QBF.

\smallskip

To increase intuitive understanding, in what follows the model $(\mathcal{M}_{\varphi}, s)$ is constructed so that the truth of $\varphi$ can be captured by a \emph{traveling game} on the model between two players: Traveler and Demon. The formula $\varphi$ is true if and only if Traveler has a winning strategy in the traveling game on $(\mathcal{M}_{\varphi}, s)$, while the \textsf{MLSR} formula $\gamma_{\varphi}$ states the existence of a winning strategy for Traveler.  $(\mathcal{M}_{\varphi}, s)$ consists of $n+1$ vertically concatenated `modules': one initial module for the first quantifier in $\varphi$, one module for each of the remaining $n-1$ quantifiers, plus one final verification module. Each of these modules is depicted in Figure \ref{Modules}. More in detail, the construction of $\mathcal{M}_{\varphi}$ is  as follows: starting with the initial module, we concatenate successive $\forall x_{i}$- and $\exists x_{j}$-modules corresponding to the order of quantifiers in $\varphi$ (we treat the top nodes labeled by $x_j$ and $\neg x_j$ as the end nodes of the previous module). The \emph{goal points} are those to which the valuation assigns the proposition letter $g$ (as depicted in Figure \ref{Modules}). Once all $n$ quantifier modules have been added, we append the final verification module. For each clause $C_{i}$, we use a distinct proposition letter $c_{i}$, which holds at exactly one node in the verification module, called a \emph{clause vertex}. Each clause vertex $c_{i}$ has an outgoing edge to all and only the duals of literals that make $C_{i}$ true.
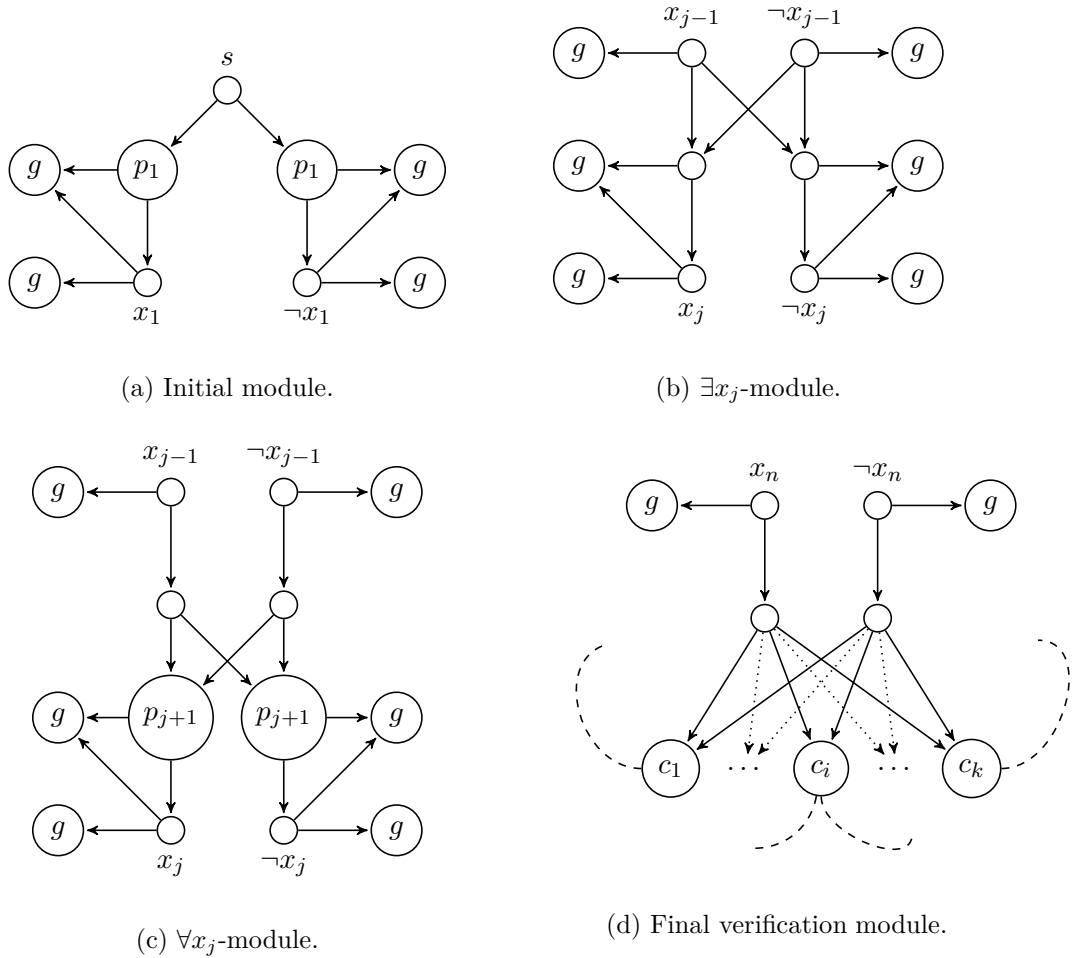
\begin{figure}[h!]
\begin{subfigure}[b]{0.55\textwidth}
\begin{center}
\begin{tikzpicture}[->,>=stealth',shorten >=1pt,auto,node distance=1.5cm,semithick]
\tikzstyle{every state}=[fill=white,draw=black,text=black, minimum size=4pt]

\node[state]         (A) [label=90:$s$]                                   {};
\node[state]         (B) [below left of=A]   {$p_1$};
\node[state]         (C) [below right of=A]  {$p_1$};
\node[state]         (D) [below of=B,    label=-90:$x_{1}$]        {};
\node[state]         (E) [below of=C,label=-90:$\neg x_{1}$]        {};
\node[state]         (F) [left of=B]         {$g$};
\node[state]         (G) [right of=C]        {$g$};
\node[state]         (H) [left of=D]         {$g$};
\node[state]         (I) [right of=E]        {$g$};

\path   (A) edge              node {} (B)
            edge              node {} (C)
        (B) edge              node {} (D)
            edge              node {} (F)
        (C) edge              node {} (E)
            edge              node {} (G)
        (D) edge              node {} (F)
            edge              node {} (H)
        (E) edge              node {} (G)
            edge              node {} (I);
\end{tikzpicture}
\end{center}
\caption{Initial module.}
\end{subfigure}
\begin{subfigure}[b]{.35\textwidth}
\begin{center}
\begin{tikzpicture}[->,>=stealth',shorten >=1pt,auto,node distance=1.5cm,semithick]
\tikzstyle{every state}=[fill=white,draw=black,text=black, minimum size=4pt]
\node[state]         (D) [label=90:$x_{j-1}$]                    {};
\node[state]         (E) [right of=D, label=90:$\neg x_{j-1}$]        {};
\node[state]         (H) [left of=D]         {$g$};
\node[state]         (I) [right of=E]        {$g$};
\node[state]         (J) [below of=D]        {};
\node[state]         (K) [below of=E]        {};
\node[state]         (L) [left of=J]         {$g$};
\node[state]         (M) [right of=K]        {$g$};
\node[state]         (N) [below of=J, label=-90:$x_{j}$]        {};
\node[state]         (O) [below of=K, label=-90:$\neg x_{j}$]        {};
\node[state]         (P) [left of=N]         {$g$};
\node[state]         (Q) [right of=O]        {$g$};

\path   (D) edge              node {} (H)
            edge              node {} (J)
            edge              node {} (K)
        (E) edge              node {} (I)
            edge              node {} (J)
            edge              node {} (K)
        (J) edge              node {} (L)
            edge              node {} (N)
        (K) edge              node {} (M)
            edge              node {} (O)
        (N) edge              node {} (L)
            edge              node {} (P)
        (O) edge              node {} (M)
            edge              node {} (Q);
\end{tikzpicture}
\end{center}
\caption{$\exists x_j$-module.}
\end{subfigure}\vspace{1em}

\begin{subfigure}{.55\textwidth}
\begin{center}
\begin{tikzpicture}[->,>=stealth',shorten >=1pt,auto,node distance=1.5cm,semithick]
\tikzstyle{every state}=[fill=white,draw=black,text=black, minimum size=4pt]
\node[state]         (N) [label=90:$x_{j-1}$]        {};
  \node[state]         (O) [right of=N, label=90:$\neg x_{j-1}$]        {};
  \node[state]         (P) [left of=N]         {$g$};
  \node[state]         (Q) [right of=O]        {$g$};
  \node[state]         (R) [below of=N]        {};
  \node[state]         (S) [below of=O]        {};
  \node[state]         (T) [below of=R]        {$p_{j+1}$};
  \node[state]         (U) [below of=S]        {$p_{j+1}$};
  \node[state]         (V) [left of=T]         {$g$};
  \node[state]         (W) [right of=U]        {$g$};
  \node[state]         (X) [below of=T, label=-90:$x_{j}$]        {};
  \node[state]         (Y) [below of=U, label=-90:$\neg x_{j}$]        {};
  \node[state]         (Z) [left of=X]         {$g$};
  \node[state]         (Z') [right of=Y]       {$g$};
  
\path   (N) edge              node {} (P) 
            edge              node {} (R)
        (O) edge              node {} (Q)
            edge              node {} (S)
        (R) edge              node {} (T)
            edge              node {} (U)
        (S) edge              node {} (T)
            edge              node {} (U)
        (T) edge              node {} (V)
            edge              node {} (X) 
        (U) edge              node {} (W)
            edge              node {} (Y)
        (X) edge              node {} (V)
            edge              node {} (Z)
        (Y) edge              node {} (W)
            edge              node {} (Z'); 
\end{tikzpicture}
\end{center}
\caption{$\forall x_j$-module.}
\end{subfigure}
\begin{subfigure}{.4\textwidth}
\begin{center}
\begin{tikzpicture}[->,>=stealth',shorten >=1pt,auto,node distance=1.5cm,semithick]
\tikzstyle{every state}=[fill=white,draw=black,text=black, minimum size=4pt]

 \node[state]         (X) [label=90:$x_{n}$]        {};
  \node[state]         (Y) [right of=X, label=90:$\neg x_{n}$]        {};
  \node[state]         (Z) [left of=X]         {$g$};
  \node[state]         (Z') [right of=Y]       {$g$};
  \node[state]         (1) [below of=X]        {};
  \node[state]         (2) [below of=Y]        {};
  \node (mid)  at ($(1)!0.5!(2)$)       {};
    \node[state]         (4) [node distance=2cm, below of=mid]  {$c_i$};
  \node[state]         (3) [node distance = 2cm, left of =4]  {$c_1$};
   \node[state]         (5) [node distance = 2cm, right of=4]  {$c_k$};
  \node (int1) at ($(3)!0.5!(4)$)       {$\dots$};
  \node (int2) at ($(4)!0.5!(5)$)       {$\dots$};
  \node (3d) [node distance=1cm,above left of=3] {};
    \node (4d) [node distance=1cm,below right of =4] {};
      \node (5d) [node distance=1cm,above right of = 5] {};
      \node (3dd) [node distance=1cm,above of = 3d] {};
    \node (4dd) [node distance=0.5cm,right of = 4d] {};
     \node (4dd2) [node distance=1.5cm,below left of = 4] {};
      \node (5dd) [node distance=1cm,above of = 5d] {};
      
       \draw[dashed,-, out=180, in=200, looseness = 1.2] (3) to (3dd);
      \draw[dashed,-,out=-90, in=-90, looseness = 1] (4) to (4dd);
      \draw[dashed,-,out=-100, in=0, looseness = 1] (4) to (4dd2);
       \draw[dashed,-,out=0, in=0, looseness = 1.2] (5) to (5dd);
       
  \draw[dotted] (1) to (int1);
  \draw[dotted] (1) to (int2);
  \draw[dotted] (2) to (int1);
  \draw[dotted] (2) to (int2);
  
 \path  (X)  edge              node {} (Z)
            edge              node {} (1)
        (Y) edge              node {} (Z')
            edge              node {} (2)
        (1) edge              node {} (3)
        (1) edge              node {} (4)
        (1) edge              node {} (5)
        (2) edge              node {} (3)
        (2) edge              node {} (4)
        (2) edge              node {} (5);

\end{tikzpicture}
\end{center}
\caption{Final verification module.}
\end{subfigure}
\caption{The shape of the initial module (a) does not depend on which quantifier $\varphi$ begins with. In (b), (c) and (d), the top nodes labeled by $x_j$ and $\neg x_j$ are the end nodes of the previous module. In (d), each clause vertex $c_{i}$ has an outgoing edge to a vertex labeled by a literal $\pm x_{j}$ exactly if the \emph{dual} literal $\mp x_{j}$ makes clause $C_{i}$ true.}
\label{Modules}
\end{figure}
\smallskip

Now, the traveling game proceeds in the following manner. At the beginning, Traveler is positioned at the starting vertex $s$. When Demon plays, she deletes a node in the graph. When Traveler plays, she can travel along one of the remaining edges to an adjacent vertex. Traveler wins if she manages to reach a goal point, marked with the proposition letter $g$. Demon wins otherwise. More in detail, if $\varphi$ starts with $\exists$, Traveler goes first. If $\varphi$ starts with $\forall$, Demon goes first: in the first move, she can only delete a vertex marked by $p_1$\textemdash that is, she can only delete a point adjacent to the starting vertex. From then on, Traveler and Demon alternate their turns, with turns being either traveling one edge further or deleting one node, respectively, where the Demon's second move at each $\forall x_{j}$-module is restricted to nodes marked with $p_{j}$ (see Figure \ref{Modules}). This continues in this manner until Traveler reaches a node that sees a clause node. At this point, Demon has $k-1$ moves, which she must use to delete all but one clause node. Then, we allow Traveler two successive moves (once Demon has restricted her choices to one clause node). Then, Demon and Traveler once again alternate single moves until the game is resolved. See Figure \ref{example} for an example.
\begin{figure}[h!]
\begin{center}
\begin{tikzpicture}[->,>=stealth',shorten >=1pt,auto,node distance=1.5cm,semithick]
  \tikzstyle{every state}=[fill=white,draw=black,text=black, minimum size=4pt]

  \node[state] (A)                     [label=90: $s$]{};
  \node (form) [node distance = 2cm, above of = A] []{$\varphi = \forall x_{1}\exists x_{2} \forall x_{3} (C_{1}\wedge C_{2} \wedge C_{3})$};
  \node (form) [node distance = 1.5cm, above of = A] []{where $C_1= \neg x_1\vee x_2$, 
$C_2= \neg x_1\vee x_2\vee\neg x_{3}$, and $C_3= x_{1} \vee x_{2} \vee  x_3$};
  \node[state]         (B) [below left of=A]   {$p_1$};
  \node[state]         (C) [below right of=A]  {$p_1$};
  \node[state]         (D) [below of=B, label=0:$x_{1}$]        {};
  \node[state]         (E) [below of=C,label=180:$\neg x_{1}$]        {};
  \node[state]         (F) [left of=B]         {$g$};
  \node[state]         (G) [right of=C]        {$g$};
  \node[state]         (H) [left of=D]         {$g$};
  \node[state]         (I) [right of=E]        {$g$};
  \node[state]         (J) [below of=D]        {};
  \node[state]         (K) [below of=E]        {};
  \node[state]         (L) [left of=J]         {$g$};
  \node[state]         (M) [right of=K]        {$g$};
  \node[state]         (N) [below of=J, label=0:$x_{2}$]        {};
  \node[state]         (O) [below of=K, label=180:$\neg x_{2}$]        {};
  \node[state]         (P) [left of=N]         {$g$};
  \node[state]         (Q) [right of=O]        {$g$};
  \node[state]         (R) [below of=N]        {};
  \node[state]         (S) [below of=O]        {};
  \node[state]         (T) [below of=R]        {$p_3$};
  \node[state]         (U) [below of=S]        {$p_3$};
  \node[state]         (V) [left of=T]         {$g$};
  \node[state]         (W) [right of=U]        {$g$};
  \node[state]         (X) [below of=T, label=0:$x_{3}$]        {};
  \node[state]         (Y) [below of=U, label=180:$\neg x_{3}$]        {};
  \node[state]         (Z) [left of=X]         {$g$};
  \node[state]         (Z') [right of=Y]       {$g$};
  \node[state]         (1) [below of=X]        {};
  \node[state]         (2) [below of=Y]        {};
  \node      (mid)  at ($(1)!0.5!(2)$)       {};
    \node[state]         (4) [node distance=2cm, below of=mid]  {$c_2$};
  \node[state]         (3) [left of = 4]  {$c_1$};
  \node[state]         (5) [right of= 4]  {$c_3$};
  \path (A) edge              node {} (B)
            edge              node {} (C)
        (B) edge              node {} (D)
            edge              node {} (F)
        (C) edge              node {} (E)
            edge              node {} (G)
        (D) edge              node {} (F)
            edge              node {} (H)
            edge              node {} (J)
            edge              node {} (K)
        (E) edge              node {} (G)
            edge              node {} (I)
            edge              node {} (J)
            edge              node {} (K)
        (J) edge              node {} (L)
            edge              node {} (N)
        (K) edge              node {} (M)
            edge              node {} (O)
        (N) edge              node {} (L)
            edge              node {} (P) 
            edge              node {} (R)
        (O) edge              node {} (M)
            edge              node {} (Q)
            edge              node {} (S)
        (R) edge              node {} (T)
            edge              node {} (U)
        (S) edge              node {} (T)
            edge              node {} (U)
        (T) edge              node {} (V)
            edge              node {} (X) 
        (U) edge              node {} (W)
            edge              node {} (Y)
        (X) edge              node {} (V)
            edge              node {} (Z)
            edge              node {} (1)
        (Y) edge              node {} (W)
            edge              node {} (Z')
            edge              node {} (2)
        (1) edge              node {} (3)
        (1) edge              node {} (4)
        (1) edge              node {} (5)
        (2) edge              node {} (3)
        (2) edge              node {} (4)
        (2) edge              node {} (5)
        (3) edge [dashed, out=180, in=200, looseness = 1.1] node {} (D)
        (3) edge [dashed, out=-40, in=-20, looseness = 1.9] node {} (O)
         (4) edge [dashed, out=210, in=200, looseness = 1.7] node {} (D)
        (4) edge [dashed, out=-40, in=-20, looseness = 1.4] node {} (O)
           (4) edge [dashed, out=210, in=-135, looseness = 2.5] node {} (X)
         (5) edge [dashed, out=0, in=-20, looseness = 1.1] node {} (E)
        (5) edge [dashed, out=35, in=-20, looseness = 1] node {} (O)
        (5) edge [dashed, out=70, in=-45, looseness = 1] node {} (Y);
\end{tikzpicture}
\end{center}\vspace{-8em}
\caption{An example. The proposition letter $g$ marks the goal points. Each $\forall$-module forces Traveler to the literal $\pm x_{j}$ point chosen by Demon, while each $\exists$-module leaves the choice to Traveler.
The letters $p_{1},...,p_{k}$ are \emph{level markers} that restrict Demon's moves. In the final module, Demon forces Traveler into some clause.
For each literal $\pm x_{j}$ that makes such clause true, Traveler can then go to the node labeled by dual literal $\mp x_{j}$ above.}\label{example}
\end{figure}
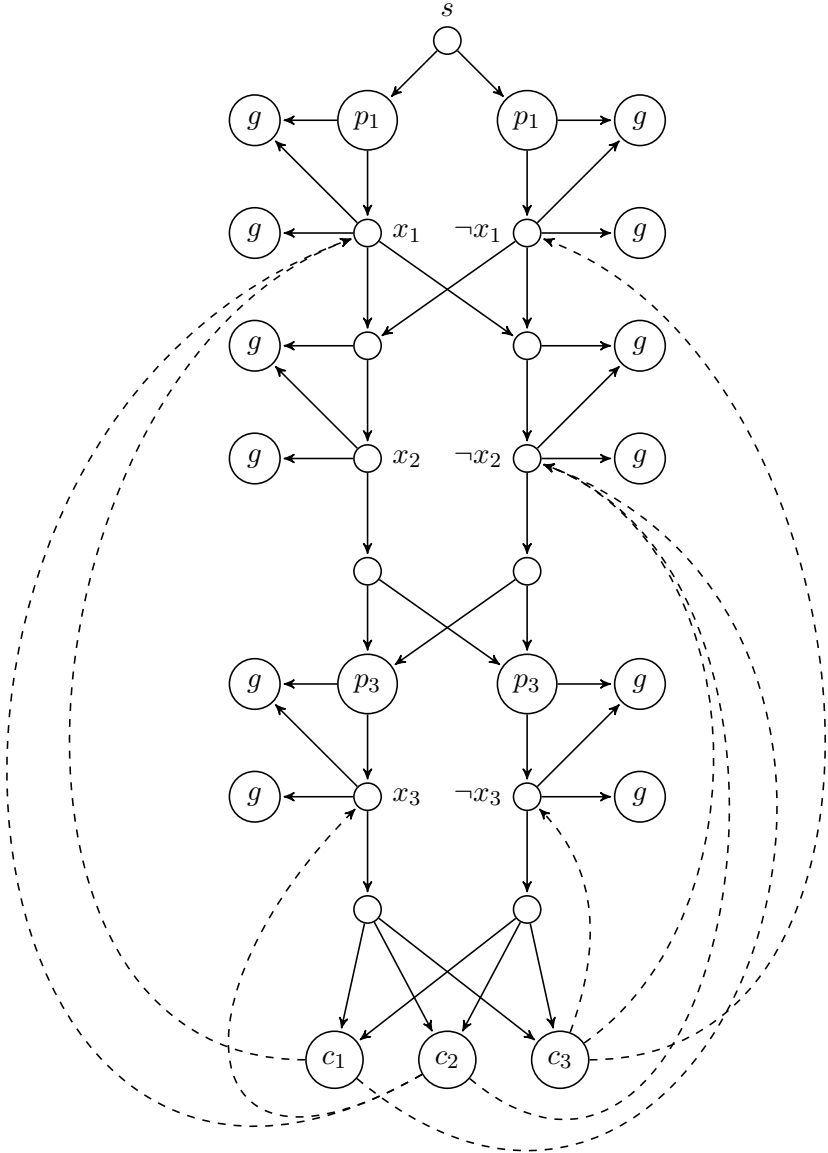

\medskip
The game adequately captures the truth of $\varphi$:

\begin{observation}
Traveler has a winning strategy for the game on $(\mathcal{M}_{\varphi}, s)$ if and only if the initial QBF formula $\varphi$ is true. 
\end{observation}

\begin{proof}
Each travel path to the verification module yields a valuation. Say that \emph{a truth value is selected for} $x_{i}$ if Traveler's path passes through the node labeled $x_{i}$. At $\forall x_{i}$ modules, Demon selects a truth value for $x_{i}$. At $\exists x_{j}$ modules, Traveler selects a truth value for $x_{j}$.  Once Traveler reaches a clause node, an assignment has thus been chosen for all variables. Here, the design of the above modules guarantees the following two key facts at that stage: (i) the deleted goal points are all and only those seen by the visited vertices, and also, (ii) all unvisited $\pm x_{i}$ vertices still have two adjacent goal points. 

If $\varphi$ is true, then the assignment chosen in this manner (with Demon controlling $\forall$ and Traveler controlling $\exists$) makes all (disjunctive) clauses true. So, for every clause $C_{i}$, there is some visited $\pm x_{j}$ node, for some $\pm x_{j}$ that entails $C_{i}$. By design of the final clause module, no matter which clause Traveler is at, there is some \emph{unvisited} vertex labeled $\mp x_{j}$ accessible from this clause vertex. Traveler can then travel to this vertex, where she sees two goal points. Demon can remove at most one of them at her next move, and Traveler therefore wins. 
Conversely, if $\varphi$ is false, the final assignment makes at least one clause false. Demon forces Traveler to the corresponding clause vertex: because the current assignment makes the disjunctive clause false, all the accessible $\pm x_{j}$ vertices have already been visited and, thus, do not see any goal points. Demon therefore wins. 
\end{proof}

Lastly, to conclude the proof of Theorem \ref{Model checking}, we make sure that \textsf{MLSR} can express the existence of a winning strategy for Traveler. When $\varphi$ starts with $\forall$ and has $n$ quantifiers and $k$ clauses, the general form of the corresponding \textsf{MLSR} formula $\gamma_{\varphi}$ is 
$$
 \big([-\alpha_{j}]\Diamond\big)^{f(n)}[\delta]^{k-1}\Diamond^{2} [-\top]\Diamond g
$$
Here, $f(n)$ is  a function counting the total number of  rounds played in the game up to the final module: $f$ is linear in $n$ (it is in fact easy to see that $f(n)\leq 3n$). The symbol $\alpha_{j}$ denotes $p_{j}$ whenever Traveler sees a $p_{j}$-point in the corresponding $\forall x_{j}$-module; it stands for $\top$ otherwise. 
The formula $\delta$, on the other hand, is a Boolean combination of $c_{i}$'s expressing that exactly one of the clauses is true. The repeated modalities capture exactly the structure of the game and the restrictions on the players' moves. The game goes on for $f(n)$ rounds until the penultimate stage is reached. The $[-p_{j}]$ modalities force Demon to remove only $p_{j}$-points during the middle round played on a $\forall x_{j}$-module. Then, $[\delta]^{k-1}$ quantifies over all ways in which Demon can remove $k-1$ clauses (all but one). The formula $\gamma_{\varphi}$ expresses that Traveler can ensure that such a sequence of moves results in reaching a goal point, and thus holds exactly if Traveler has a winning strategy: equivalently, it holds if and only if the initial QBF formula $\varphi$ is true.\footnote{The results in this section extend to the expanded modal language of \S\ref{Axiomatization}. The PSPACE lower bound obviously remains valid, but so is the upper bound. The reason is that extending the first-order translation of Fact \ref{translation} to nominals and \textsf{PAL}-modalities incurs only  a polynomial blow-up in size.}
\end{proof}

\noindent\textbf{Note on game perspectives.} While not strictly necessary for the proof of Theorem \ref{completeness thm}, the above traveling game with point removal over the initial structure is independently appealing, and it suggests links with the graph games that motivate sabotage logics and related logics for graph change mentioned in \S\ref{Model change and quantification} \citep{vBL:2018}. As a further perspective, the above traveling game is virtually identical to the standard logical evaluation game for the crucial quantified Boolean formula in the above proof. Making these game perspectives precise is left as an open problem here.

\subsection{Satisfiability}

Next, we show that, despite the recursive axiomatizability shown in \S\ref{Axiomatization}, stepwise removal has a complex theory. The satisfiability problem for the logic \textsf{MLSR} is undecidable, which we establish by a reduction from the tiling problem, a standard technique in modal logic (cf. \citep{BRV:2001, Marx:2006}, to which we refer for details).

\begin{thm}
The satisfiability problem for \emph{\textsf{MLSR}} with two binary accessibility relations $R_u$ and $R_r$ is undecidable.
\end{thm}
\begin{proof}
Let $\mathcal{T}= \{T_1, ..., T_n\}$ be a finite set of tile types. Given a tile type $T_i$, $u(T_i), r(T_i)$, $d(T_i)$ and $l(T_i)$ will represent the colors of the upper, right, lower and left edges of $T_i$, respectively. For each tile type $T_i$, we fix a proposition letter $t_i$ that is going to encode $T_i$. We will now define an \textsf{MLSR} formula $\varphi_\mathcal{T}$ such that the following holds:
$$\varphi_\mathcal{T} \text{ is satisfiable if and only if } \mathcal{T} \text{ tiles the discrete quadrant } \mathbb{N}\times\mathbb{N}.
$$
\noindent The formula $\varphi_\mathcal{T}$ is the conjunction of the following \textsf{MLSR} formulas. The first three describe the relational structure of a grid, the last three encode the behavior of a tiling of the grid:

\begin{align*}
(\emph{Func}) &&& \mathsf{U}\langle-\top \rangle (\Box_{u}\bot \wedge \Diamond_{r}\top)    \\ &&& \mathsf{U}\langle-\top \rangle (\Box_{r}\bot \wedge \Diamond_{u}\top)   \\
(\emph{Conf}) &&& \mathsf{U}\langle-\top \rangle (\Diamond_{r}\Box_{u}\bot \wedge \Diamond_{u}\Box_{r}\bot ) \\
(\emph{Unique}) &&& \mathsf{U}\Bigg(\bigvee_{1\leq i\leq n} t_i\wedge\bigwedge_{1\leq i<j\leq n} (t_i\rightarrow\neg t_j)\Bigg)\\
(\emph{Vert}) &&& \mathsf{U}\bigwedge_{1\leq i\leq n}\Bigg(t_i\rightarrow\Diamond_u\bigvee_{1\leq j\leq n, u(T_i) = d(T_j)} t_j\Bigg)\\
(\emph{Horiz}) &&& \mathsf{U}\bigwedge_{1\leq i\leq n}\Bigg(t_i\rightarrow \Diamond_r\bigvee_{1\leq j\leq n, r(T_i) = l(T_j)} t_j\Bigg)
\end{align*}

\smallskip
\noindent ($\Leftarrow$) It is easy to see that any tiling of 
$\mathbb{N}\times\mathbb{N}$ induces a model for $\varphi_\mathcal{T}$.

\smallskip

\noindent ($\Rightarrow$) For the other direction, suppose that $\mathcal{M}, w\models\varphi_\mathcal{T}$, for some $\mathsf{LSR}$-model $\mathcal{M}=(W, R_u, R_r,$ $V)$ and $w\in W$. The formula (\emph{Func}) ensures that the relations $R_{u}$ and $R_r$ are functions, and that for every point $x$, that $R_{u}[x]\neq R_{r}[x]$ (the $R_{u}$ and $R_{r}$-images of $x$ are different). The formula (\emph{Conf}) then guarantees that the functions commute: $R_{u}\circ R_{r} = R_{r}\circ R_{u}$. This ensures the existence of an embedding $f:\mathbb{N}^{2}\rightarrow W$ that preserves the structure of vertical and horizontal successors: that is, for all $(n,m)\in\mathbb{N}^{2}$, we have $R_{u}(f(n,m), f(n,m+1))$ and $R_{r}(f(n,m), f(n+1,m))$. Now, tile the point $(n,m)$ in $\mathbb{N}^{2}$ with tile $T_{i}$ exactly if $\mathcal{M}, f(n,m) \models t_{i}$. This gives a tiling of the discrete quadrant of the plane.
\end{proof} 

The two standard modalities used in this proof can be reduced to one using standard techniques \citep{KW:1999}, but we forego details here because of the syntactic cost involved in writing the formulas.

The above undecidability argument applies a fortiori to the richer language of \S\ref{Axiomatization} with nominals and \textsf{PAL} modlities. But it will also work with languages that are less expressive than \textsf{MLSR}. In particular, one can replace the universal modality by an extra standard modality that can survey the domain by employing the well-known `spypoint technique' from hybrid logic. A detailed syntactic construction of this sort for modal logics of graph games can be found in \citep{ZBMA:2019}.

Having determined the complexity of model checking and satisfiability, one task would remain, concerning definability  and expressive power. However, we leave this open here.

\begin{ope*}
What is the complexity of testing for \emph{\textsf{SR}}-bisimulation?
\end{ope*}

Given any two finite models $\mathcal{M}$, $\mathcal{N}$, it is easy to find an EXPSPACE upper bound. One considers the space of all models arising from $\mathcal{M}$, $\mathcal{N}$ by deleting finite sequences of different points, and then tests for ordinary modal bisimulation over this space with respect to \textsf{MLSR}, now viewed as a standard bimodal language. But, just as with standard modal bisimulation \citep{KS:1983}, one can probably do better.

\section{Stepwise removal over first-order fragments}\label{Stepwise removal over first-order fragments}

Having established the complexity of adding quantification without replacement to the basic modal language, we can also consider other fragments of first-order logic. Perhaps the simplest case is adding the removal modality $\del{\varphi}\psi$ to monadic first-order logic \textsf{MFOL}. As it turns out, this yields exactly the formulas  in \textsf{MFOL}$_{=}^{x}$: that is, all formulas with one free variable $x$ in monadic first-order logic with identity.

\begin{thm}\label{MFOLx=}
$\mathsf{MLSR(MFOL)}=\mathsf{MFOL}_{=}^{x}$.
\end{thm}
\begin{proof}
Fix finitely many unary predicates $P_{1},\dots, P_{k}$.  We define standard normal forms for the whole language $\mathsf{MFOL_{=}}$. \emph{Local state descriptions} $sd$ are conjunctions of $\pm P_{i}$ with $1\leq i \leq k$. There are $2^{k}$ of these, and they can be applied to arbitrary variables. \emph{Global state descriptions} $SD$ of depth $N$ are then conjunctions $\bigwedge_{j} SD_{j}$ where, for each local state description $sd_{j}$, $SD_{j}$ is either the statement that exactly $m_{j}$ objects satisfy $sd_{j}$, where we have $m_{j}<N$, or the statement that at least $N$ objects satisfy $sd_{j}$. 

\begin{defn} An \emph{enumerative normal form} is a disjunction of conjunctions $NF$, each consisting of (a) local state descriptions for each of the variables $x_{i}$ plus a complete set of equalities and inequalities for all pairs of variables from $x_{1}, ..., x_{m}$, plus (b) a global state description that is consistent with (a) in an obvious syntactic sense.
\end{defn}

\begin{cla*}
Each formula in $\mathsf{MFOL_{=}}$ of quantifier depth $N$ and $m$ free variables $x_{1}, ..., x_{m} $ is equivalent to  an enumerative normal form. 
\end{cla*}

\noindent This can be proved by induction on formulas via a syntactic argument.\footnote{Alternatively, $NF$ describes a model $\mathcal{M}$ in such a way that, for any model $\mathcal{N}$ that satisfies $NF$, Duplicator has a winning strategy in the Ehrenfeucht game over $N$ rounds between $\mathcal{M}$ and $\mathcal{N}$ starting with the partial isomorphism between the objects on both sides satisfying the atomic diagram (a).} 

\begin{cla*}
$\mathsf{MFOL_{=}}$ is closed under the modality $\del{\varphi}\psi$.    
\end{cla*}
\noindent \emph{Proof of Claim.} Using the disjunction axioms for existential modalities and $\del{\varphi}\psi$ stated in \S\ref{Axiomatization}, in proving closure, one can restrict attention to conjunctive forms $NF$ and special removal modalities $\del{sd\wedge SD}NF$. Closure can be shown here by a simple argument, driven by the following two key facts:

\begin{itemize}
    \item the equivalence $\del{sd\wedge SD}NF\leftrightarrow (SD\wedge \del{sd}NF)$ is valid,\footnote{Here is a general useful principle that is easy to state in first-order syntax. When we take out a point satisfying $\varphi(x)\wedge\psi$, where $x$ does not occur free in $\psi$, then we can just put $\psi$ outside in a conjunction.}
    \item the formula $\del{sd_{i}}(sd' \wedge SD$) is equivalent to $sd' \wedge SD[i:=i+1]$,
\end{itemize}
where $SD[i:=i+1]$ replaces the quantification in the $i$-th conjunct of $SD$ by a quantifier stating the existence of one more point satisfying the relevant local state description.        
\end{proof}

Arguments like this are available for other languages that admit of simple normal forms of modal depth 1. Here is one obvious question. 
\begin{ope*}
What fragment of first-order logic results from adding the dynamic operators of $\mathsf{MLSR}$ to the language of modal $\mathsf{S5}$?
\end{ope*}

We have some initial results, but the combinatorics get considerably more complex, since the logic can now also distinguish between different equivalence classes in $\mathsf{S5}$ models.

\smallskip
The general question suggested by the specific case analyzed in Theorem \ref{MFOLx=} is the following: what is the boost in expressive power when we close fragments of first-order logic under various model-changing modalities?

\section{Conclusion and further directions}\label{Conclusion}

The logic of stepwise removal of objects lies in between  modal logics of definable model change and
 logics for graph games with arbitrary moves, and it may well be the most intuitive example of a modal system that crosses the line from decidable to undecidable.\footnote{Another contender is the `modal fact change logic' of \cite{Declan:2019}.} We have established its main properties, proving a bisimulation characterization theorem and other results on expressive power, a completeness theorem, and two basic complexity results. Most of the techniques that we used are well-known, others less so, and we also introduced a new technique for proving completeness. The resulting style of thinking can be applied to a wide range of modal systems of this sort.

Among the issues still to be addressed is the complexity problem for \textsf{SR}-bisimulation (\S\ref{Complexity and undecidability}), as well as the expressive closure problem for $\mathsf{S5}$. Moreover, all of our questions return for some obvious extensions and variations.

\medskip

\noindent\textbf{Simultaneous versions of \textsf{MLSR}.} It is natural to add a modality for removing  a fixed finite number of points, either in a conjunctive unary version $\langle-(\varphi_1,...,\varphi_k)\rangle\psi$ or with truly polyadic operators $\langle-\overline{\varphi}\rangle\psi$, where the formulas $\overline{\varphi}$ can be evaluated in a tuple of indices. These modalities seem undefinable as iterations of our unary $\langle-\varphi\rangle\psi$. Even so, we conjecture that all of our results go through.

\medskip

Another immediate question concerns other extended modal logics.

\medskip

\noindent \textbf{Connections with hybrid logic.} \textsf{MLSR} seems closely related to hybrid modal formalisms such as `memory logics' that have been used to detect jumps to undecidability for fragments of \textsf{}{FOL} in an illuminating manner, \citep{AFFM:2008}. Given that \textsf{MLSR} translates into a fragment of the first-order language, cf. Fact \ref{not LSR-def}, it may be of interest to compare the fragments that arise by adding the removal modality $\langle-\overline{\varphi}\rangle\psi$ to various first-order fragments with the natural hierarchy offered by the memory-logic perspective.\footnote{The referee has suggested looking also at a variant of \textsf{MLSR}, where $\langle-\overline{\varphi}\rangle\psi$ only describes taking away a point satisfying $\psi$ that is $R$-accessible from the current point. This modified logic of `accessible' object removal translates into $H(@,\downarrow)$, i.e., the \emph{bounded  fragment} of first-order logic.}

\medskip

Then, there is the question of the scope of our methods.

\medskip

\noindent\textbf{Axiomatizing logics of graph games.} It is a long-standing open problem how to axiomatize the validities of sabotage-style modal logics and related ones \citep{AvBG:2018}. Does our axiomatization technique for \textsf{MLSR} employing added dynamic epistemic modalities work for these logics, as well?\footnote{Modifications may be needed: e.g., for sabotage logics, one wants to name arrows rather than points.}
\medskip

Next, returning to the issue of undecidability, a few questions arise naturally. 

\medskip
\noindent\textbf{Other sources of undecidability.} In addition to the undecidability induced by stepwise removal, there is the undecidability induced by local link-cutting or local definable point removal, taking place only at the current point of evaluation \citep{Dazhu:2018}. Both modifications of dynamic-epistemic logics block the usual recursion axioms, both allow for tiling encodings, but the connection remains to be clarified.

\medskip
But there are also other perspectives on complexity that we have found.

\medskip

\noindent\textbf{Lowering the complexity of \textsf{MLSR}.} Can \textsf{MLSR} be shifted back into the decidable modal fold? For many logical systems, one can lower the complexity by a Henkin-style change in the semantics \citep{ABBN:2016}. In particular, one could restrict the removal of points to those that are accessible from the current point in some global relation $Axy$ and, if this does not suffice for decidability, one might use further guarding, so that the earlier first-order translations of \textsf{MLSR} formulas (from \S\ref{Basics of expressive power}) end up inside guarded, or loosely guarded, fragments of \textsf{FOL}.

Yet, moves like this make most sense when connected to a principled view of computation. We believe that modal logics like \textsf{MLSR}, but also hybrid memory logics or related systems,  offer an interesting alternative take on the sources of computational complexity. In the usual automata hierarchy, Turing machine power arises when we have an active memory that can be rewritten. In our logics, however, a simple device that merely stores the set of deleted or visited points suffices. The reason must be the interplay of memory and expressiveness of the language for constructing models around that memory, suggesting a sort of descriptive complexity theory complementary to that of \cite{IM:1999}.

\medskip
\medskip
\noindent {\bf Acknowledgments} \; We thank audiences in Amsterdam, Beijing, Gothenburg and Stanford for their comments on versions of this work, while Dazhu Li and the referee provided further useful corrections. We thank, especially, Alexandru Baltag for giving us the crucial hint toward our completeness theorem. But most of all, we are indebted to Carlos Areces for a very inspiring and pleasant collaboration during his 2018 spring stay at Stanford. Johan van Benthem was supported by the Tsinghua University Initiative Scientific Research Program, No. 2017THZWYX08.


\begin{thebibliography}{24}
\providecommand{\natexlab}[1]{#1}
\providecommand{\url}[1]{\texttt{#1}}
\expandafter\ifx\csname urlstyle\endcsname\relax
  \providecommand{\doi}[1]{doi: #1}\else
  \providecommand{\doi}{doi: \begingroup \urlstyle{rm}\Url}\fi

\bibitem[Andr\'{e}ka et~al.(2016)Andr\'{e}ka, Bezhanishvili, van Benthem, and
  N\'{e}meti]{ABBN:2016}
H.~Andr\'{e}ka, N.~Bezhanishvili, J.~van Benthem, and I.~N\'{e}meti.
\newblock Changing a {S}emantics: {O}pportunism or {C}ourage?
\newblock In E.~Alonso, M.~Manzano, and I.~Sain, editors, \emph{The Life and
  Work of Leon Henkin}, pages 307--337. Birkhaueser Verlag, 2016.

\bibitem[Areces and ten Cate(2006)]{AtC:2006}
C.~Areces and B.~ten Cate.
\newblock Hybrid {L}ogics.
\newblock In P.~Blackburn, J.~van Benthem, and F.~Wolter, editors,
  \emph{Handbook of {M}odal {L}ogic}, pages 821--868. Elsevier Science,
  Amsterdam, 2006.

\bibitem[Areces et~al.(2008)Areces, Figueira, Figueira, and Mera]{AFFM:2008}
C.~Areces, D.~Figueira, S.~Figueira, and S.~Mera.
\newblock Expressive {P}ower and {D}ecidability for {M}emory {L}ogics.
\newblock In W.~Hodges and R.~de~Queiroz, editors, \emph{Logic, Language,
  Information and Computation, WoLLIC 2008. Lecture Notes in Computer Science},
  volume 5110, pages 56--68. Springer, Berlin, Heidelberg, 2008.

\bibitem[Areces et~al.(2015)Areces, Fervari, and Hoffmann]{AFH:2015}
C.~Areces, R.~Fervari, and G.~Hoffmann.
\newblock Relation-{C}hanging {M}odal {O}perators.
\newblock \emph{Logic Journal of the IGPL}, 23\penalty0 (4):\penalty0 601--627,
  2015.

\bibitem[Aucher et~al.(2018)Aucher, van Benthem, and Grossi]{AvBG:2018}
G.~Aucher, J.~van Benthem, and D.~Grossi.
\newblock Modal logics of sabotage revisited.
\newblock \emph{Journal of Logic and Computation}, 28\penalty0 (2):\penalty0
  \,269--303, 2018.

\bibitem[{\VAN{Benthem}{van}{van}}~Benthem(2005)]{vB:2005}
J.~{\VAN{Benthem}{van}{van}}~Benthem.
\newblock An essay on sabotage and obstruction.
\newblock \emph{\emph{In D. Hutter and W. Stephan (eds.),} Mechanizing
  Mathematical Reasoning: Essays in Honor of J\"{o}rg H. Siekmann on the
  Occasion of His 60th Birthday, Lecture Notes in Artificial Intelligence},
  2605:\penalty0 \,268--276, 2005.

\bibitem[{\VAN{Benthem}{van}{van}}~Benthem(2011)]{JB:2011}
J.~{\VAN{Benthem}{van}{van}}~Benthem.
\newblock \emph{Logical {D}ynamics of {I}nformation and {I}nteraction}.
\newblock Cambridge University Press, Cambridge UK, 2011.

\bibitem[{\VAN{Benthem}{van}{van}}~Benthem and Liu(2020)]{vBL:2018}
J.~{\VAN{Benthem}{van}{van}}~Benthem and F.~Liu.
\newblock Graph {G}ames and {L}ogic {D}esign.
\newblock In F.~Liu, H.~Ono, and J.~Yu, editors, \emph{Knowledge, Proof and
  Dynamics}, pages 125--146. Logic in Asia: Studia Logica Library, Springer:
  Singapore, 2020.

\bibitem[{\VAN{Benthem}{van}{van}}~Benthem
  et~al.(2009){\VAN{Benthem}{van}{van}}~Benthem, Gerbrandy, Hoshi, and
  Pacuit]{BGHP:2009}
J.~{\VAN{Benthem}{van}{van}}~Benthem, J.~Gerbrandy, T.~Hoshi, and E.~Pacuit.
\newblock Merging frameworks for interaction.
\newblock \emph{Journal of Philosophical Logic}, 38\penalty0 (5):\penalty0 \,
  491--526, 2009.

\bibitem[Blackburn et~al.(2011)Blackburn, de~Rijke, and Venema]{BRV:2001}
P.~Blackburn, M.~de~Rijke, and Y.~Venema.
\newblock \emph{Modal {L}ogic}.
\newblock Cambridge University Press, Cambridge UK, 2011.

\bibitem[Gabbay(2013)]{DG:2013}
D.~Gabbay.
\newblock Introducing {R}eactive {K}ripke {S}emantics and {A}rc
  {A}ccessibility.
\newblock In D.~Gabbay, editor, \emph{Reactive {K}ripke {S}emantics}, pages
  29--76. Springer Science, Dordrecht, 2013.

\bibitem[Hansen(2011)]{JUH:2011}
J.~U. Hansen.
\newblock A hybrid public announcement logic with distributed knowledge.
\newblock \emph{Electronic Notes in Theoretical Computer Science, Elsevier
  Amsterdam}, 273:\penalty0 \,33--50, 2011.

\bibitem[Hintikka and Sandu(1997)]{HS:1997}
J.~Hintikka and G.~Sandu.
\newblock Game-{T}heoretic {S}emantics.
\newblock In A.~{\VAN{Meulen}{ter}{ter}}~Meulen and J.~van Benthem, editors,
  \emph{Handbook of {L}ogic and {L}anguage}, pages 361--410. Elsevier Science,
  Amsterdam, 1997.

\bibitem[Immerman(1999)]{IM:1999}
N.~Immerman.
\newblock \emph{Descriptive {C}omplexity}.
\newblock Springer Science Publishers, Dordrecht, 1999.

\bibitem[Kanellakis and Smolka(1983)]{KS:1983}
P.~Kanellakis and S.~Smolka.
\newblock {CCS} expressions, finite state processes, and three problems of
  equivalence.
\newblock In \emph{Proceedings of the 2nd ACM Symposium on Principles of
  Distributed Computing}, pages 228--240. Springer Science, Dordrecht, 1983.

\bibitem[Kracht and Wolter(1999)]{KW:1999}
M.~Kracht and F.~Wolter.
\newblock Normal {M}onomodal {L}ogics can {S}imulate {A}ll {O}thers.
\newblock \emph{Journal of Symbolic Logic}, 64\penalty0 (1):\penalty0 99--138,
  1999.

\bibitem[Li(2020)]{Dazhu:2018}
D.~Li.
\newblock Losing {C}onnection: {T}he {M}odal {L}ogic of {D}efinable {L}ink
  {D}eletion.
\newblock \emph{Journal of Logic and Computation}, 30\penalty0 (3):\penalty0
  715--743, 2020.

\bibitem[L\"{o}ding and Rohde(2003)]{LR:2003}
C.~L\"{o}ding and P.~Rohde.
\newblock Solving the {S}abotage {G}ame is {PSPACE}-{H}ard.
\newblock \emph{\emph{In Rovan B., Vojt\'{a}s P. (eds.),} Mathematical
  Foundations of Computer Science 2003. MFCS 2003. Lecture Notes in Computer
  Science}, 2747, 2003.

\bibitem[Marx(2006)]{Marx:2006}
M.~Marx.
\newblock Complexity of {M}odal {L}ogic.
\newblock In P.~Blackburn, J.~van Benthem, and F.~Wolter, editors,
  \emph{Handbook of {M}odal {L}ogic}, pages 139--179. Elsevier Science,
  Amsterdam, 2006.

\bibitem[Renardel~de Lavalette(2001)]{REN:2001}
G.~Renardel~de Lavalette.
\newblock A {L}ogic of {M}odification and {C}reation.
\newblock In C.~Condoravdi and G.~Renardel~de Lavalette, editors, \emph{Logical
  {P}erspectives on {L}anguage and {I}nformation}, pages 197--219. CSLI
  Publications, Stanford, 2001.

\bibitem[Rohde(2005)]{Roh:2005}
P.~Rohde.
\newblock On games and logics over dynamically changing structures.
\newblock \emph{\emph{Ph.D. dissertation, RWTH Aachen University, Germany}},
  pages 1--216, 2005.

\bibitem[Stockmeyer and Meyer(1973)]{SM:1973}
L.~J. Stockmeyer and A.~R. Meyer.
\newblock Word problems requiring exponential time.
\newblock \emph{Proceedings of the 5th ACM Symposium on Theory of Computing,
  STOC '73}, pages \,1--9, 1973.

\bibitem[Thompson(2020)]{Declan:2019}
D.~Thompson.
\newblock Local {F}act {C}hange {L}ogic.
\newblock In F.~Liu, H.~Ono, and J.~Yu, editors, \emph{Knowledge, Proof and
  Dynamics}, pages 73--96. Logic in Asia: Studia Logica Library, Springer:
  Singapore, 2020.

\bibitem[Zaffora~Blando et~al.(2020)Zaffora~Blando, Mierzewski, and
  Areces]{ZBMA:2019}
F.~Zaffora~Blando, K.~Mierzewski, and C.~Areces.
\newblock The {M}odal {L}ogics of the {P}oison {G}ame.
\newblock In F.~Liu, H.~Ono, and J.~Yu, editors, \emph{Knowledge, Proof and
  Dynamics}, pages 3--23. Logic in Asia: Studia Logica Library, Springer:
  Singapore, 2020.

\end{thebibliography}
\end{document}